\documentclass[%
 reprint,
 amsmath,amssymb,
 aps,
]{revtex4-2}

\usepackage{graphicx}
\usepackage{dcolumn}
\usepackage{bm}
\usepackage{float}
\usepackage{tabularx}
\usepackage{setspace}
\usepackage{hyperref}
\usepackage{outlines}
\usepackage{bbm}

\usepackage{xcolor}
\newcommand{\editcolor}[0]{black}

\usepackage{xr}
\makeatletter
\newcommand*{\addFileDependency}[1]{
  \typeout{(#1)}
  \@addtofilelist{#1}
  \IfFileExists{#1}{}{\typeout{No file #1.}}
}
\makeatother
\newcommand*{\myexternaldocument}[1]{%
    \externaldocument{#1}%
    \addFileDependency{#1.tex}%
    \addFileDependency{#1.aux}%
}
\myexternaldocument{SI}
\usepackage{appendix}

\newcommand{\bl}{{\boldsymbol{\lambda}}}
\newcommand{\bL}{{\boldsymbol{\Lambda}}}

\newcommand{\bp}{\boldsymbol{\pi}}

\newcommand{\brho}{\boldsymbol{\rho}}
\newcommand{\bt}{\boldsymbol{\theta}}

\begin{document}

\singlespacing

\title{Local imperfect feedback control in non-equilibrium biophysical systems \\ enabled by thermodynamic constraints}

\author{Carlos Floyd}
\email{csfloyd@uchicago.edu}
\affiliation{The Chicago Center for Theoretical Chemistry, The University of Chicago, Chicago, Illinois 60637, USA
}
\affiliation{Department of Chemistry, The University of Chicago, Chicago, Illinois 60637, USA
}
\affiliation{The James Franck Institute, The University of Chicago, Chicago, Illinois 60637, USA
}
\author{Aaron R.\ Dinner}
\affiliation{The Chicago Center for Theoretical Chemistry, The University of Chicago, Chicago, Illinois 60637, USA
}
\affiliation{Department of Chemistry, The University of Chicago, Chicago, Illinois 60637, USA
}
\affiliation{The James Franck Institute, The University of Chicago, Chicago, Illinois 60637, USA
}
\author{Suriyanarayanan Vaikuntanathan}
\email{svaikunt@uchicago.edu}
\affiliation{The Chicago Center for Theoretical Chemistry, The University of Chicago, Chicago, Illinois 60637, USA
}
\affiliation{Department of Chemistry, The University of Chicago, Chicago, Illinois 60637, USA
}
\affiliation{The James Franck Institute, The University of Chicago, Chicago, Illinois 60637, USA
}

\date{\today}

\begin{abstract}
\color{\editcolor}
How biological networks achieve robust control despite relying on imperfect, local information remains an important open question.  Here, we identify thermodynamic constraints that can curtail non-equilibrium steady-state responses so severely that even crude, local feedback rules can achieve globally stable control without requiring precise network design or global information. Specifically, using Markov jump processes as a general framework for biophysical dynamics, we derive general non-equilibrium response constraints showing that for many classes of rate perturbations, steady-state responses have fixed signs across all driving strengths, so that near-equilibrium responses predict far-from-equilibrium behavior regardless of system complexity. These constraints clarify several biological phenomena: monotonicity is thermodynamically guaranteed whenever a perturbation acts on a single transition rate, and non-monotonic responses, as observed for example in transcription factor regulation, arise only when an input simultaneously modulates multiple rates. Even in this case, we identify a graph-theoretic concept termed ``coherence'' that allows for a restoration of monotonicity. We show how coherence naturally and generally emerges in classic biophysical models of adaptation, including \textit{E. coli} chemotaxis, and transcription factor regulation when biological constraints on network parameterization are included. We next show that, within a control-theoretic framework, these constraints guarantee that simple linear feedback on small subsets of kinetic rates achieves globally stable tracking and adaptation without coordinated manipulation of many variables. For systems with one regulator, local stability implies global stability for arbitrary network topologies without fine tuning.  For multiple regulators, the thermodynamic constraints restrict the allowed nullcline geometry in ways that preclude unstable dynamical structures.  We also show that stable control with non-monotonic responses can be achieved by tuning the signs of different feedback couplings for the regulated variable.  Together, these results reveal that non-equilibrium thermodynamics fundamentally constrains biochemical network responses, explaining when and why responses are monotonic or non-monotonic and guaranteeing the effectiveness of minimal, local feedback for robust control in complex reaction networks.
\color{black}
\end{abstract}

\maketitle

\begin{twocolumngrid}    

\begin{figure*}
\centering
\includegraphics[width=\textwidth]{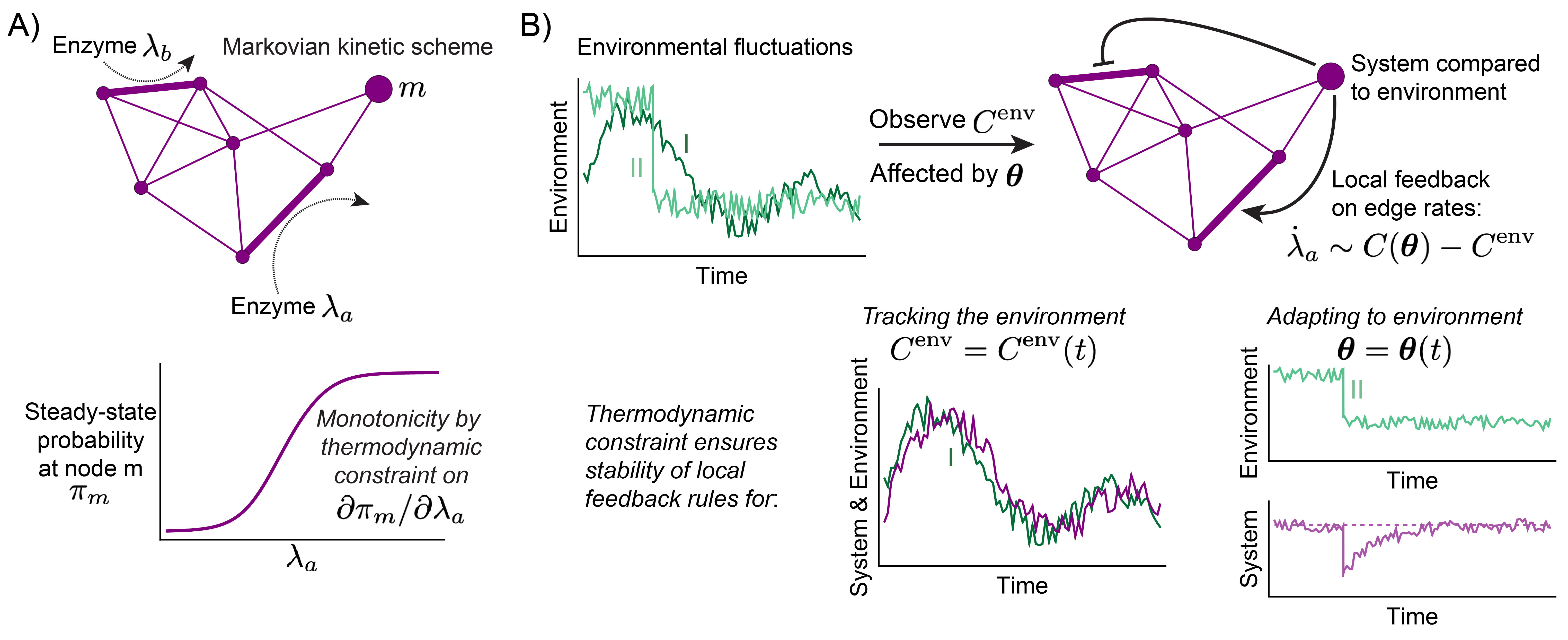}
\caption{\textbf{A thermodynamic constraint on non-equilibrium response enables control of Markov jump processes by local feedback connections.}  A)  The response of the non-equilibrium steady-state probability $\pi_m$ at node $m$ to a change in a rate perturbation $\lambda_a$ is strictly monotonic.  This manifests as a partial derivative $\partial \pi_m / \partial \lambda_a$ having a fixed sign across the range of $\lambda_a$.  B)  This monotonicity property enables dynamical control using local, linear feedback loops even from initial conditions far away from the desired fixed points.  We consider a system-environment decomposition in which fluctuating environmental variables enter as external functions in the feedback for driving forces or in the network parameters.  Depending on the nature of the feedback, the system can track the environmental variables or adapt to perturbations by returning to a set point.}
\label{NewFig1}
\end{figure*}

\section{Introduction}

A common challenge in learning, training, or adaptation tasks is reliance on imperfect or incomplete information about the system. Despite this difficulty, there are several notable examples where such information is used successfully. For instance, the adaptive immune system must rely on imperfect spatiotemporally local information, such as estimates of binding affinities between antigens and receptors on cells \cite{mayer2019well}. These and other biological responses nonetheless appear to make near-optimal use of available information, even in time-varying environments \cite{mayer2019well, mattingly2021escherichia, schnaack2022learning, iram2021controlling, murugan2021roadmap}. How imperfect local information can guide control in complex environments remains an important and open question.

A key difficulty in answering this question is that many biological control and adaptation tasks occur in non-equilibrium settings, where there are fewer thermodynamic guarantees on stability compared to equilibrium settings \cite{lan2012energy, tu2018adaptation,mayer2019well, landmann2021simple}.  Furthermore, given the high dimensionality and complexity of biological systems, one might expect that different initial conditions would lead to different (meta)stable states, i.e., local attractors, making optimization sensitive to the precise mechanism of feedback control. Here, we prove a general thermodynamic constraint on the response of steady states to changes in driving forces for a broad class of non-equilibrium processes.  Specifically, we show that a steady state’s partial derivative with respect to the driving force maintains a fixed sign for all values of the force, which we refer to as a monotonicity property.  This is akin to the fixed-sign properties of partial derivatives of free energy that provide thermodynamic stability in equilibrium settings \cite{callen1991thermodynamics}. As a consequence of this constraint, the solution space for complex control problems can be surprisingly simple, often featuring a global attractor.

\color{\editcolor}
We establish this constraint using a graph-theoretic proof within the context of Markov jump processes, a modeling framework encompassing many biophysical processes (Figures \ref{NewFig1} and \ref{MonotonicityProof}).  Next, we outline a systematic criterion for establishing monotonicity of non-equilibrium responses even when multiple parameters are regulated. We refer to these as the coherence conditions.  We then leverage these properties of monotonicity to show how non-equilibrium systems can effectively mount dynamical responses to time-dependent environmental changes using minimal, linear feedback connections between observables defined on node probabilities and non-equilibrium driving forces. We discuss how such linear feedback rules can occur and be useful in biological contexts including \textit{E. coli} chemotaxis and transcription factor networks with condensate or sequestration-mediated feedback. 
\color{black}
Indeed, they can be viewed as integral feedback control rules which, due to our thermodynamic constraints, are effective for a wide range of topologies and interactions without the need for fine tuning or engineering~\cite{ma2009defining}. In line with the concept of imperfect gradients used in machine learning \cite{lillicrap2016random, lillicrap2020backpropagation, wright2022deep}, these linear feedback connections serve as local and fixed approximations to globally stable feedback mechanisms which would require complete (non-local) and instantaneous information about the network state.

\color{black}
\begin{figure*}
\centering
\includegraphics[width=\textwidth]{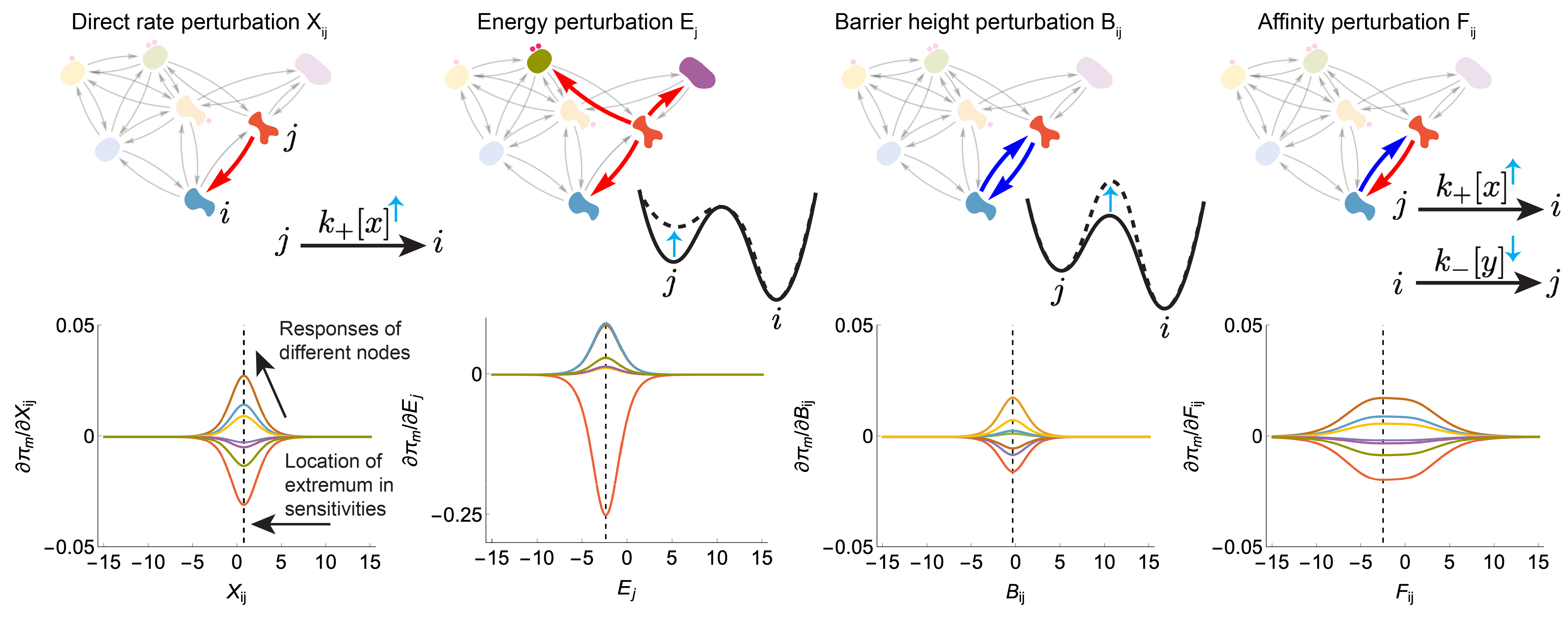}
\caption{
\color{\editcolor}
\textbf{Non-equilibrium responses to several rate perturbations are strictly monotonic.} For a random graph of first-order transitions between molecular states, we show several types of rate perturbations corresponding to the parameterization in Equation~\ref{eqmarkovrate}. Red arrows indicate directed edges effectively increased by the parameter and blue arrows indicate edges decreased by it. The bottom row shows plots of the steady-state derivatives for randomly sampled base edge rate values, with lines colored according to the node $m$ they correspond to in the graph. The dashed vertical line marks the value of parameter, generically denoted by $\lambda$ following the convention in the main text, at which the derivatives of all nodes are simultaneously extremized.
\color{black}
}
\label{MonotonicityProof}
\end{figure*}

We demonstrate the utility of these simple feedback connections in two tasks: one where the system updates an internal model of the environmental state and another where it adapts to an environmental perturbation by returning to a set point.  Importantly, for systems with one or two chemical regulators, our monotonicity constraint implies the global stability of these minimal feedback connections \textit{for arbitrary network topology}; no specific motifs or modular designs are necessary \cite{alon2019introduction}. Our results, which are enabled by the thermodynamic constraints we derive, can be viewed as simple, easily implementable, and local learning rules that can be used to track changing environmental conditions.  Such guarantees could, in application, facilitate the design of next-generation synthetic biochemical feedback systems \cite{nandagopal2011synthetic, chen2024synthetic, del2015biomolecular, araujo2018topological, araujo2023universal, ma2009defining, aoki2019universal, PRXLife.3.013017, briat2016antithetic}.

\section{Modeling non-equilibrium molecular processes with rate perturbations}

In many biological and physical systems, sustained external forcing keeps the system away from equilibrium. A central question is how the steady-state probabilities change as a function of various kinetic parameters and applied driving forces \cite{owen2020universal, owen2023size, aslyamov2024general, aslyamov2024nonequilibrium, aslyamov2025nonequilibrium,fernandes2024topologically, liang2024thermodynamic, harunari2024mutual}. Here we model systems as Markov jump processes over discrete sets of $N_\text{n}$ states.  
The state of the system is described by the probability vector $\mathbf{p}(t) \in \mathbb{R}^{N_\text{n}}$ which evolves according to $\dot{\mathbf{p}} = \mathbf{W}(\bl(t);\bt) \mathbf{p}$, where $\bl(t) \in \mathbb{R}^{D}$ represents a possibly time-dependent control vector (described below) and $\bt$ represents the default kinetic rates.  Under the assumption of ergodicity, for a fixed $\bl$ the system relaxes to a unique steady state $\bp(\bl;\bt) = \lim_{t\to\infty} \mathbf{p}(t)$.  Observables $C$ that are functions of the state, such as the weighted average occupancy of a set of nodes, can be computed as 
\begin{equation}
    C(\mathbf{p}) = \sum_{m} q_m p_m, \label{eqCdefOrig}
\end{equation} 
with $q_m$ denoting the weight of node $m$.

\color{\editcolor}
Throughout this paper, we focus primarily on unimolecular (i.e., linear or first-order) reaction networks. In this setting, the stochastic description over discrete states becomes equivalent in the large copy number (mean-field) limit to deterministic rate equations for the corresponding state concentrations, allowing the probability vector $\mathbf{p}(t)$ to be interpreted as a concentration vector. We will incorporate feedback through a control vector $\bl$ that modulates selected first-order transition rates in $\mathbf{W}$. In the macroscopic limit this allows for deterministic dynamics $\dot{\bl}(\mathbf{p}(t))$ that depend on the instantaneous set of concentrations. Although bimolecular networks likewise reduce to deterministic nonlinear mass-action equations at large copy number, their underlying microscopic state spaces grow combinatorially, and the accompanying nonlinear coupling between transitions makes edge-specific control less transparent.  While our analytical stability results therefore apply rigorously to mean-field unimolecular processes, we later discuss how related monotonicity constraints and local feedback constructions may extend to bimolecular systems.

Transitions between states are modeled as Poisson processes, with the rate from state $j$ to state $i$ given by the rate matrix element $W_{ij}$. In common biological settings, such as transcriptional regulation, enzymatic control of phosphorylation, or voltage-gating of membrane channels, feedback variables may modulate these rates through different localized or collective ways. To capture this flexibility, we adopt the following intentionally over-parameterized representation:
\begin{equation}
W_{ij} = e^{X_{ij}} e^{E_j - B_{ij} + F_{ij}/2}. \label{eqmarkovrate}
\end{equation}
Here $X_{ij}$ is a transition-specific rate factor; $E_j$ is a state-dependent energy that affects all transitions out of state $j$; $B_{ij} = B_{ji}$ is a symmetric barrier height influencing both directions of the $i \leftrightarrow j$ edge; and $F_{ij} = -F_{ji}$ represents antisymmetric non-conservative driving forces, such as chemical potential differences  
\footnote{One may also split $F_{ij}$ asymmetrically among $W_{ij}\propto e^{fF_{ij}}$ and $W_{ji}\propto e^{(1-f)F_{ij}}$ for any $f \in (0,1)$.  We consider $f=1/2$ throughout here.}.  
These parameters provide a convenient representation of the collective perturbations that biological systems may impart to a chemical process, rather than a unique physical factorization of the rate matrix. The vector ${\bm \lambda}$ is used to denote this set of perturbations. 
We illustrate these parameter perturbations in Figure \ref{MonotonicityProof}.  Equation \ref{eqmarkovrate} defines a smooth, unique mapping from kinetic coordinates to rate matrices, and feedback operates by perturbing these coordinates to induce corresponding changes in the rates. Because we proceed forward from parameters to rates, inversion of this mapping is not required. In what follows, components $\lambda_a$ of the control vector are used to modulate selected elements of this parameter set, enabling analysis of general conditions for stable feedback control across different types of rate perturbations.

\begin{figure*}[ht!]
\begin{center}
\includegraphics[width=\textwidth]{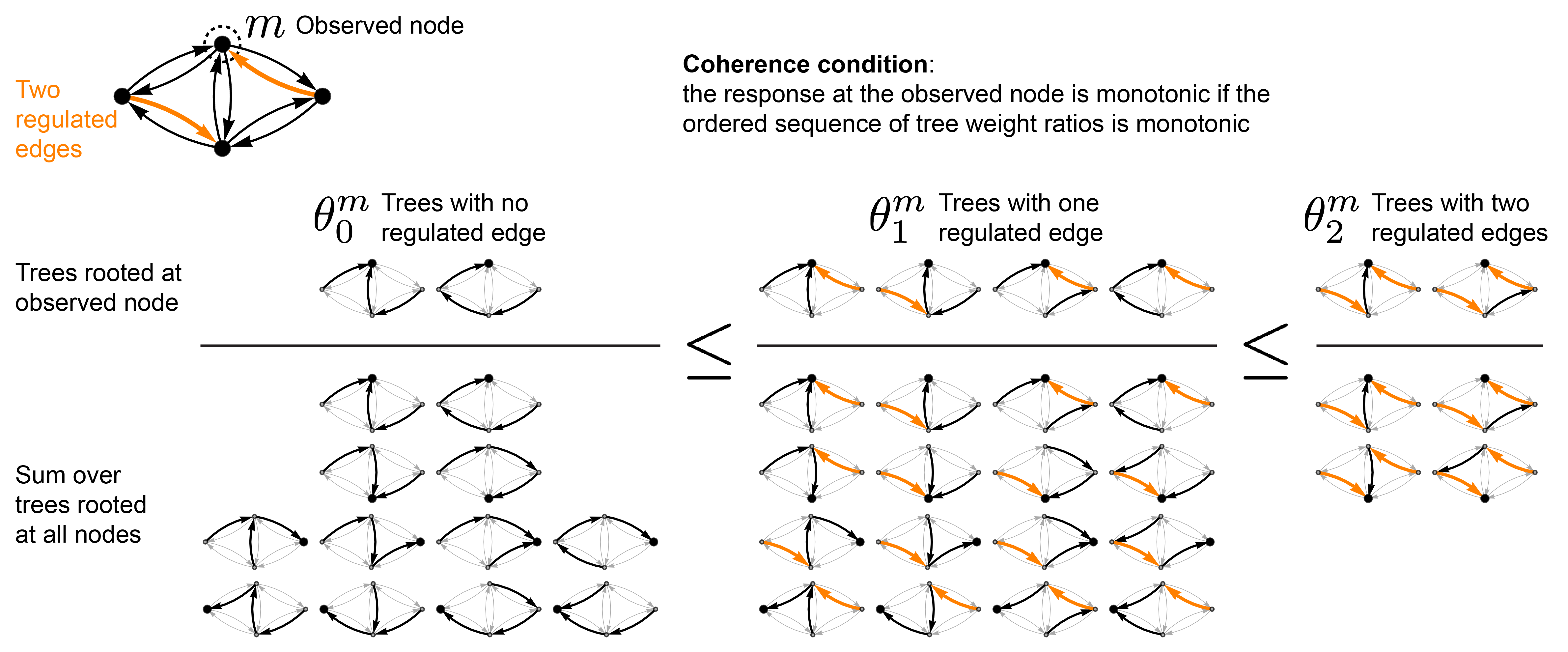}
\caption{
\color{\editcolor}
\textbf{Example of positive coherence in a four-state network.} Consider a network with two transitions (highlighted in orange) directly regulated by $\lambda$. We examine the monotonicity of $\pi_m(\lambda)$, the steady-state probability at node $m$ (circled), as a function of $\lambda$. The bottom of the figure shows the spanning tree decomposition underlying the construction of the ratios $\theta_0^m,\ \theta_1^m,$ and $\theta_2^m$, which group trees according to how many regulated edges they contain. In each ratio, the numerator sums over trees rooted at $m$, while the denominator sums over trees rooted at all nodes (marked by black circular symbols), providing the normalization. In both numerator and denominator, each tree contributes its tree weight, defined as the product of the edge rates $W_{ij}$ along its edges with $\lambda$ factored out.  
\color{black}}
\label{CoherenceDef}
\end{center}
\end{figure*}

\section{Thermodynamic constraints on far-from-equilibrium responses}\label{sec:ThermoConstraints}
\subsection{General mathematical framework}
Our central result is the derivation of thermodynamic constraints on the non-equilibrium responses, i.e., the derivative $\partial C / \partial \lambda$ which quantifies how the steady-state observable $C(\bp(\lambda))$ changes as a function of various rate perturbations $\lambda$. For a broad class of perturbation types and observables, we prove that when all other parameters in the network are held fixed, this derivative maintains a fixed sign across the range of $\lambda$. In other words, the observable $C$ responds strictly monotonically to changes in $\lambda$. This fixed sign property of a non-equilibrium response can be viewed as analogous to the fixed sign of free-energy second derivatives in equilibrium thermodynamics \cite{callen1991thermodynamics}. 

We first describe a general, sufficient condition for the response of a steady-state node probability $\pi_m(\lambda)$ to a change in some parameter $\lambda > 0$ to be monotonic.  We apply the matrix-tree theorem \cite{schnakenberg1976network, hill2013free}, which exactly gives the steady-state probability $\pi_m$ at an arbitrary node $m$ in terms of a sum over directed spanning trees of the Markov network, with each tree given a weight equal to the product of the transition rates along its directed edges.   Application of this graph-theoretic construction for $\pi_m(\lambda)$ yields rational polynomial functions of $\lambda$ of the form \cite{floyd2025limits}
\begin{equation}
    \pi_m(\lambda) = 
    \frac{\sum_{\mu=0}^{N_\lambda} \zeta^m_\mu \lambda^\mu}
         {\sum_{\nu=0}^{N_\lambda} \bar{\zeta}_\nu \lambda^\nu}
    \equiv 
    \sum_{\mu=0}^{N_\lambda} 
    w_\mu(\lambda)\,\theta_\mu^m , \label{eqpimlambda}
\end{equation}
where $\bar{\zeta}_\mu = \sum_k \zeta^k_\mu$ is the sum over all nodes $k$ of $\zeta_\mu^k$, which are the sums of tree weights for all directed trees rooted at $k$ and containing $\mu$ powers of contribution from the parameter $\lambda$.  $N_\lambda$ is the number of unique algebraic powers of $\lambda$ among all such tree weights, and both denominator and numerator are written the same maximum power $N_\lambda$ provided coefficients are allowed to be zero. Because the denominator represents a normalization, the highest power in the numerator can never exceed that in the denominator.  We have introduced the weights
\begin{equation}
    w_\mu(\lambda) = 
    \frac{\bar{\zeta}_\mu \lambda^\mu}
         {\sum_{\nu=0}^{N_\lambda} \bar{\zeta}_\nu \lambda^\nu},
\end{equation} 
and the ratios 
\begin{equation}
    \theta_\mu^m = \frac{\zeta^m_\mu}{\bar{\zeta}_\mu}. \label{eqthetadef}
\end{equation}
These ratios quantify the relative contributions of tree weights with $\mu$ powers of $\lambda$ flowing into node $m$ compared to those flowing into all nodes.  

In Appendix \ref{App:CoherenceProofs} we show that a sufficient condition for Equation \ref{eqpimlambda} to be monotonic in $\lambda$ is that either
\begin{equation}
    \theta_0^m \leq \theta_1^m \leq \cdots \leq 
    \theta_{N_\lambda}^m , \label{eqmonorderinc}
\end{equation}
in which case $\pi_m(\lambda)$ increases monotonically, or
\begin{equation}
    \theta_0^m \geq \theta_1^m \geq \cdots \geq 
    \theta_{N_\lambda}^m, \label{eqmonorderdec}
\end{equation}
in which case $\pi_m(\lambda)$ decreases monotonically.  We refer to these as the ``coherence conditions.''   If either of these two orderings is broken, then $\pi_m(\lambda)$ may vary non-monotonically in $\lambda$.  Intuitively, these conditions state that trees with increasing powers $\mu$ of contributions from $\lambda$ must increasingly (or decreasingly) direct flow to node $m$ relative to other nodes.  We illustrate the graph construction behind these conditions in a four-state network example in Figure \ref{CoherenceDef}.

\textit{Responses are always monotonic when $N_\lambda=1$.}  An important special case is $N_\lambda = 1$, for which the response is monotonic because one of the two orderings applies automatically.  In Appendix \ref{App:CoherenceProofs} we also show that, in this case, the responses $\partial \pi_m/ \partial \lambda$ and $\partial \pi_{m'}/\partial \lambda$ for two different states are proportional to each other, satisfying 
\begin{equation}
    \frac{\partial \pi_m / \partial \lambda}{\partial \pi_{m'} / \partial \lambda} = R_{m,m'} \label{eqRmm}
\end{equation}
with $R_{m,m'}$ independent of $\lambda$. 

Additionally, for $N_\lambda = 1$ one can show that the derivatives  $\partial \pi_m / \partial \Lambda$ are each extremized at $\Lambda = \ln(\bar{\zeta}_0 / \bar{\zeta}_1)$ and symmetric about  this extremum, where we define 
\begin{equation}
    \Lambda \equiv \ln \lambda.
\end{equation} 
We use $\Lambda \in \mathbb{R}$ and $\lambda >0$ as regulated variables interchangeably throughout the paper, choosing whichever is more natural for visualization or exposition.  As these variables are related by a monotonic mapping, monotonicity in one variable is equivalent to monotonicity in the other.

\textit{Coherence conditions ensure monotonicity when tuning $X_{ij}$, $E_j$, or $B_{ij}$ because $N_\lambda = 1$.} We now examine how the coherence conditions manifest for specific rate perturbations, with $\lambda$ corresponding to changes in $X_{ij}$ for a chosen transition $j \rightarrow i$, $B_{ij}$ or $F_{ij}$ for a chosen transition pair $j \rightarrow i$ and $i \rightarrow j$, or $E_j$ for some chosen state $j$.  Unless stated otherwise, indices $i$ and $j$ refer throughout to edges and nodes under regulation by $\lambda$.  As shown in Appendix \ref{App:SpecificPerturbations}, when $\lambda$ corresponds to changes in $X_{ij}$, $E_j$, or $B_{ij}$, Equation \ref{eqpimlambda} has $N_\lambda = 1$ so the non-equilibrium response to these perturbations is strictly monotonic and proportional across all nodes. 

\textit{Coherence conditions ensure monotonicity when tuning $F_{ij}$ despite $N_\lambda = 2$.} Surprisingly, even when $\lambda$ affects $F_{ij}$, in which case $N_\lambda = 2$, the coherence conditions continue to hold, ensuring that the responses remain both monotonic and proportional, although the responses are no longer symmetric about the extremum.  That the coherence conditions hold when $\lambda$ scales $F_{ij}$ is due to non-trivial tree-level constraints among the $\theta_\mu^m$ ratios that we derive; see Appendix \ref{App:SpecificPerturbations} and Figures~\ref{SI_TreeSwapping} and \ref{SI_MonNumerics}.  Together, these results, illustrated in Figure \ref{MonotonicityProof}, demonstrate that for several biologically relevant classes of rate perturbations, the steady-state occupancies respond in a constrained manner, highlighting a robust structural property of non-equilibrium networks.

\textit{Input multiplicity $M$ constrains monotonicity.}  More generally, we consider how the coherence conditions are modified when a single control variable $\lambda$ simultaneously perturbs multiple rate parameters. In Ref.~\citenum{floyd2025limits} we showed that this ``input multiplicity'' $M$, i.e., the number of rate parameters affected by a given control variable, can exponentially enhance the information-processing capacity of reaction networks, and in Ref.~\citenum{owen2023size} it is shown that this multiplicity can bound the sharpness of responses. In the present context, achieving non-monotonic non-equilibrium responses requires $M>1$, since only then can the powers $N_\lambda$ in Equation~\ref{eqpimlambda} exceed their baseline values (see Appendix \ref{App:SpecificPerturbations}). For higher $M$ the non-equilibrium responses are not generally proportional to each other.  However, breaking monotonicity requires specific relationships among the coefficients $\theta_\mu^m$ that violate the coherence conditions. We discuss this further below with two biological examples (Sections ~\ref{sbsec:Adap} and\ref{sbsec:TF}) and show how known biochemical limitations on network topology can restore coherence even when multiple rates are tuned simultaneously.  

\textit{Monotonicity of any non-equilibrium average observable guaranteed when $M=1$.}  Finally, we study the non-equilibrium response of collective network observables rather than individual node occupancies. Any observable defined as a positive linear combination $C = \sum_m q_m \pi_m$ with $q_m > 0$, retains the structural form of Equation~\ref{eqpimlambda}, with the coefficients $q_m$ entering into the corresponding $\theta^m_\mu$ terms:
\begin{equation}
    C = \sum_{\mu=0}^{N_\lambda} w_\mu(\lambda)\sum_m q_m \theta^m_\mu \equiv \sum_{\mu=0}^{N_\lambda} w_\mu(\lambda)\theta^C_\mu, \label{eqCdef}
\end{equation}
where 
\begin{equation}
    \theta^C_\mu = \frac{\sum_m q_m \zeta_\mu^m}{\bar{\zeta}_\mu} \equiv \frac{\zeta^C_\mu}{\bar{\zeta}_\mu}. \label{eqthetaCdef}
\end{equation}
In this case, the coherence conditions depend not only on the kinetic rates $W_{ij}$ but also on the choice of observable itself through the ratios $\theta^C_\mu$. Here due to Equation~\ref{eqRmm} when $M=1$ the response is guaranteed to be monotonic for any $C(\lambda)$ and choice of rate perturbation, including $F_{ij}$ for which $N_\lambda = 2$.  

When $M=2$, then whether $C(\lambda)$ satisfies the coherence conditions and is therefore monotonic depends both on the network rates as well as on the choice of weights which define the observable $C$. To summarize, when $M=1$, the responses of any averaged non-equilibrium steady state observable are guaranteed to be monotonic. For $M>1$, this monotonicity is dictated by a modified coherence condition. Below, we show how this modified coherence condition, while analytically not as tractable as before, seems to be generically satisfied in important example biologically relevant cases.

\begin{figure*}
\begin{center}
\includegraphics[width=\textwidth]{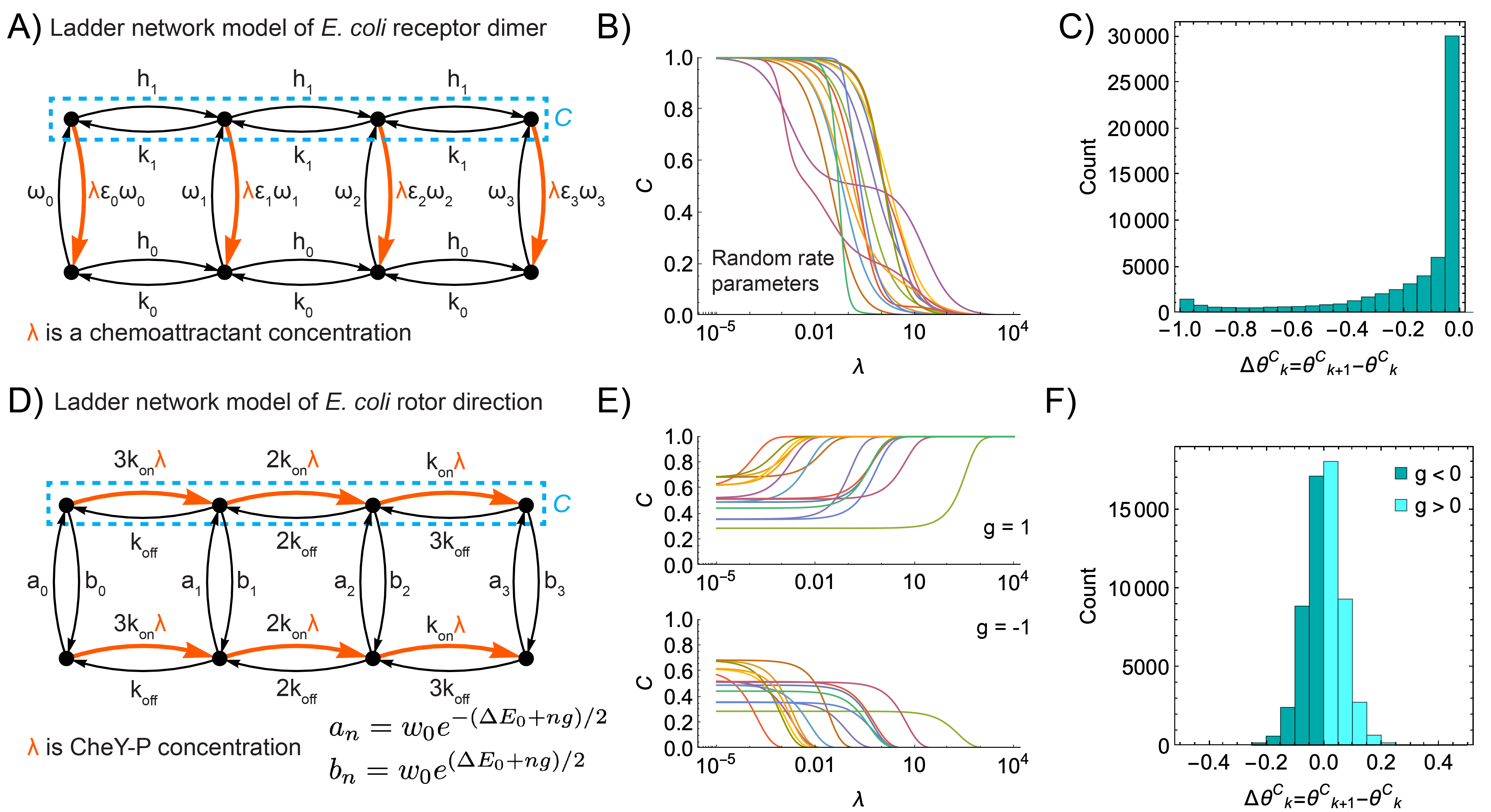}
\caption{
\color{\editcolor}
\textbf{Monotonicity despite high input multiplicity in topologically ordered models.}   
A) Ladder network for \textit{E. coli} chemoreceptor dimer activation and methylation dynamics, following Ref.~\citenum{lan2012energy}. The rate parameters $\omega_n$, $\epsilon_n$, $h_0$, $k_0$, $h_1$, and $k_1$ are labeled for $n = 0,\ldots,N$ and ladder length $N=3$. The observable $C$ sums occupancy over active dimer states (top row), and chemoattractant concentration enters through $\lambda$, which scales the deactivating edges.
B) Steady-state response $C(\lambda)$ for $N=14$ over 15 random parameter samples, each drawn log-uniformly from $[10^{-2}, 10^2]$.
C) Histogram of pairwise increments in the ratios $\theta^C_k$ (c.f. Equation~\ref{eqthetaCdef}). We drew $10^4$ parameter samples from the same distribution as panel B for $N=5$ and computed increments for each sample.
D) Same as panel C, but for the rotor direction model of Ref.~\citenum{bai2012coupling}. Rate parameters $a_n$, $b_n$, $k_{\text{off}}$, $k_{\text{on}}$ are labeled for $n=0,\ldots,N$ and $N=3$. Vertical rates follow $a_n = w_0 e^{-(\Delta E_0 + ng)/2}$ and $b_n = w_0 e^{(\Delta E_0 + ng)/2}$, with $\lambda$ scaling the rightward edges via CheY-P.
E) Same as panel B for the rotor model with $N=14$. The rate parameters $k_{\text{on}}$, $k_{\text{off}}$, $w_0$ are drawn log-uniformly from $[10^{-2},10^2]$, $\Delta E_0$ is drawn uniformly from $[-1,1]$, and we set $g=1$ (top) and $g=-1$ (bottom).
F) Same as panel C for the rotor model, with parameters drawn as in panel E and $g$ sampled uniformly from $[-1,1]$. The $10^4$ samples are split according to the sign of the sampled $g$.
\color{black}
}
\label{Ladders}
\end{center}
\end{figure*}

\subsection{Illustration of coherence in models of \textit{E. coli} chemotaxis: guarantees of monotonicity and stabilization of adaptation}
\label{sbsec:Adap}
To illustrate the use of the coherence condition framework, we next present numerical data for a classic biological adaptive system, \textit{E. coli} chemotaxis, in which monotonic non-equilibrium responses are maintained even when the input multiplicity $M$ is arbitrarily high \cite{barkai1997robustness,owen2023size,murugan2017topologically}.  This is due to ordered topologies of the system kinetics which preserve the coherence conditions at any value of $M$.  We first study the model in Ref.~\citenum{lan2012energy} shown in Figure~\ref{Ladders}A, in which an adaptive ladder-like network for the methylation (horizontal) and activation (vertical) states of a chemoreceptor dimer evolves under input from a chemoattractant that accelerates the downward deactivating transitions through pseudo-first-order rate modulation, i.e., a $\exp X_{ij}$-like perturbation. In this model, the input multiplicity for $\lambda$, the chemoattractant concentration, increases linearly with the length of the ladder (corresponding to the number of methylation states). Surprisingly, we find that for this choice of topology and rates, coherence-like conditions emerge organically and ensure that the observable $C(\lambda)$ that sums the occupancy of the active dimer states is always monotonic with respect to $\lambda$. 

In Figure~\ref{Ladders}B we sample random rate parameters for long ladder models with  15 methylation states (with 15 chosen as an arbitrary large number) and observe that monotonicity of $C(\lambda)$ is preserved in each case. We then sample $10^4$ rate parameter sets for models with $6$ methylation states and compute the ratios $\theta^C_k$ (c.f. Equation \ref{eqthetaCdef}) for the observable $C$ from the rational polynomial coefficients of the steady-state vector. From these we compute the increments $\Delta \theta^C_k = \theta^C_{k+1} - \theta^C_k$ for each power $k$. The histogram of these increments across all samples is shown in Figure~\ref{Ladders}C and indicates that each increment is negative. Thus the coherence conditions in Equation~\ref{eqmonorderdec} hold for all sampled networks.

We next study the model in Ref.~\citenum{bai2012coupling}, shown in Figure~\ref{Ladders}D, which describes a ladder-like network for binding of CheY-P to identical rotor subunits in the flagellar motor of \textit{E. coli}. The top row corresponds to the counter-clockwise rotor state, whose nodes define the observable $C$, and the bottom row corresponds to the clockwise state. A free energy gradient parameterized by $g$ drives cyclic flux within each square of the ladder. The CheY-P concentration scales the rightward transitions through a $\exp X_{ij}$-like perturbation on both rows, corresponding to binding of successive CheY-P molecules. The input $\lambda$ therefore has a multiplicity that scales as twice the ladder length. 

Despite this high input multiplicity, we observe that the non-equilibrium responses $C(\lambda)$ remain monotonic for ladders with $15$ rotor subunits when $g$ is either positive or negative (Figure~\ref{Ladders}E). For $g=1$, $C(\lambda)$ is always monotonically increasing, while for $g=-1$ it is always monotonically decreasing. We then sample $10^4$ parameter sets for models with $6$ rotor subunits and compute the coherence ratio increments $\Delta \theta^C_k$ for each case. These increments separate into strictly positive or strictly negative sets depending on the sign of $g$ (Figure~\ref{Ladders}F), demonstrating coherent responses across all sampled networks and a direct link between response sign and the free energy gradient.  

These numerical results show that, despite having large kinetic schemes, classic biological adaptive systems such as the \textit{E. coli} receptor dimer and flagellar motor exhibit strongly constrained non-equilibrium responses. The regular topologies and structured parameterization of these networks appear to enforce this behavior. We leave a detailed analytical understanding of this constraint for future work, but emphasize that results here demonstrate that the coherence conditions in Equations~\ref{eqmonorderinc} and \ref{eqmonorderdec} provide a practical and quantitative criterion for probing monotonicity. In the subsequent section, we show how this monotonicity can enable simple control schemes that, in turn, can be a basis for stable adaptation. In effect, previous requirements for adaptation, such as those in ~\cite{lan2012energy}, are recast in more general terms as coherence or monotonicity conditions. 

\subsection{Illustration of coherence in transcription regulatory models}
\label{sbsec:TF}
Recent illustrations of non-monotonic responses arising from rate modulation in models of transcriptional regulation are presented in Refs.~\citenum{mahdavi2024flexibility} and \citenum{martinez2025emergence}. These works adopt a common modeling framework, schematically shown in Figure~\ref{BiologicalExamples}A, in which gene regulation by transcription factor (TF) binding is described using a Markov jump process. In the parameters of Equation \ref{eqmarkovrate}, modulation of transition rates through TF binding appears as $\exp X_{ij}$ terms, reflecting changes in pseudo-first-order rate constants due to chemostatted TF concentrations $[\text{TF}]$. Because TF binding modifies two directed edges in the Markov graph, we have $M = N_\lambda = 2$ for such systems, and monotonicity of the response is therefore not guaranteed.

In Ref.~\citenum{martinez2025emergence}, the emergence of non-monotonic dependence of a weighted network observable on TF concentration is attributed to the presence of multiple ``incoherent'' pathways by which TF binding influences the observable. In Appendix~\ref{App:Incoherence}, we show that the specific sources of incoherence identified in that work correspond to violations of the general coherence conditions given in Equations~\ref{eqmonorderinc} and \ref{eqmonorderdec}.  The interpretation of the coherence conditions for this system is that the signs of the effects of TF on RNAP binding, RNAP unbinding, and RNAP-mediated transcription of DNA all work in the same direction to increase or decrease the transcription rate. 

In the next section, we use this monotonicity property to design and analyze control strategies for non-equilibrium systems. The fixed-sign structure of the response suggests that even systems with complex first-order transition networks can track environmental changes using simple local rules. 

\color{black}
\section{Thermodynamic constraints enable simple rules for tracking and adaptation}

Having established thermodynamic constraints on the monotonicity of the non-equilibrium response $C(\lambda)$, we now illustrate how this property enables Markov jump processes to exhibit robust tracking and adaptation behaviors via local imperfect feedback loops. We consider two generic scenarios (see Figure \ref{NewFig1}): one in which a network observable tracks a time-dependent environmental signal, and another in which environmental perturbations alter the network rates, necessitating homeostatic maintenance of a network observable at a set point.

While devising general control strategies for non-equilibrium chemical dynamics is inherently challenging \cite{yi2000robust, del2015biomolecular, araujo2018topological, araujo2023universal, ma2009defining, aoki2019universal, PRXLife.3.013017, briat2016antithetic, ilker2022shortcuts}, we show that a simple local rule can suffice for effective control in a wide range of settings.  In this rule we update the edge perturbation $\Lambda_a = \ln \lambda_a$ according to
\begin{equation}
\dot{\Lambda}_a = S_{a1} \left(C_1(\mathbf{p}) - C_1^{\text{env}}\right), \label{eq1Ddyn}
\end{equation}
where $S_{a1}$ is a fixed scalar constant, $C_1(\mathbf{p})$ denotes the current value of a network observable, and $C_1^\text{env}$ is its environmentally determined target value.  We refer to Equation~\ref{eq1Ddyn} as local because only one network observable, rather than the whole network state, is used to provide feedback to $\lambda_a$, and it is imperfect in the sense that it only approximates a universally stable feedback rule based on a Lyapunov function (see Appendix \ref{App:FormFeed}).  Since $\lambda_a$ is monotonically related to $\Lambda_a$, one may equivalently write the feedback in terms of $\lambda_a$ with the same sign, provided the dynamics do not drive $\lambda_a$ to nonphysical negative values.  Depending on the sign of $S_{a1}$, the term proportional to $C_1^\text{env}$ represents either a degradation or production term for the variable $\Lambda_a$, while the term proportional to $C_1(\mathbf{p})$ represents a process with the opposite effect. We emphasize that Equation \ref{eq1Ddyn} is linear in a single observable $C_1(\mathbf{p})$. While this simple linear form might be expected to achieve control close to a fixed point, we show in the subsequent sections that the thermodynamic constraint derived above can enable this linear form to achieve control and adaptation globally. 

\color{\editcolor}
In Appendices~\ref{App:FormFeed} and \ref{App:Non-adia} we explore the feedback dynamics outside the adiabatic limit in which  $\mathbf{p}(t) = \bp(\bL(t))$. In Appendix~\ref{App:FormFeed} we illustrate how, in this regime, the feedback gain $S_{a1}$ must not greatly exceed the internal relaxation timescales of the network, as too large a gain produces unstable oscillations analogous to those seen in other feedback control systems \cite{bechhoefer2021control, del2015biomolecular}. In Appendix~\ref{App:Non-adia} we derive the first-order correction to the adiabatic approximation for the closed-loop dynamics, which provides partial analytical insight but does not make the onset of instability fully tractable, since the perturbative expansion typically breaks down before the loss of dynamical stability. A more complete characterization of the critical gain, and in particular its dependence on the topology of the Markov graph and the choice of regulated edges, remains an open problem.
\color{black}

\textit{Biological realizations of local feedback dynamics}.  Such feedback loops as in Equation \ref{eq1Ddyn} can be readily achieved for example in transcriptional circuits, where gene products can interfere with the TFs that bind to promoter regions~\cite{alon2019introduction, reyer2021kinetic}. 
\color{\editcolor}  As a function of TF occupancy, transcription of DNA through RNA polymerase (RNAP) activity then produces a gene product (GP) such as RNA or its translated proteins, and these can then regulate transcription rates through various mechanisms \cite{parisutham2025coli}.  
\color{black}  In the well-known lac operon system, a similar biological machinery is used to track environmental concentrations of lactose and glucose and switch between states of digesting lactose or not \cite{vilar2003modeling}.  The classic model for this system includes an environmental input of lactose concentration whose metabolite allolactose de-activates a repressor for the lac operon suite of genes.  In the lac operon example, $C^{\text{env}}$ can be thought to represent the concentration of allolactose. As we demonstrate in the subsequent sections, the feedback rule and our thermodynamic constraint then ensure that the amount of GP naturally track the amount of allolactose stably and reliably, without needing further fine tuning.  The GP in this case codes for proteins which can metabolize and transport the environmentally available lactose into the cell.  Our thermodynamic constraint can hence potentially explain how the transcriptional activity of the lac operon can dynamically shift to match lactose metabolism capabilities to the current environmental availability of lactose, with feedback loop including allolactose-induced activation of transcription.
\color{\editcolor}
Finally, we note feedback rules of the sort in Equation~\ref{eq1Ddyn} along with the results of Figure~\ref{Ladders} can help ensure stable adaptation in \textit{E. coli} chemotaxis. Previous works have implicated energy consumption and other structural requirements such as features of the receptor or motor activation responses \cite{barkai1997robustness,lan2012energy}; our analysis above justifies this assumption through application of the coherence conditions (Figure \ref{Ladders}). Viewed through the lens of our results, these previous requirements ensure coherence and monotonicity, thus providing the required conditions for stable adaptive responses.
\color{black}

\color{\editcolor}
More broadly, the feedback structure of Equation~\ref{eq1Ddyn} is common in biology. In Ref.~\cite{Russo2025softmodes} bacterial responses due to stress from amino acid starvation, carbon limitation, and temperature are all viewed in an integral control framework similar to that in Equation~\ref{eq1Ddyn}. In this case, the concentration of (p)ppGpp is used to modulate transcription, translation, and replication rates across the genome~\cite{Traxler2008,Wu2022}. The eukaryotic cAMP pathway operates similarly: glucose, amino acid, nitrogen, and osmotic signals are integrated into a single cAMP concentration that coordinates downstream regulation through PKA~\cite{Zaman2008,Conrad2014}. The yeast heat shock response provides another example, with Hsf1 integrating proteotoxic stress and feeding back through Hsp70-mediated titration~\cite{Zheng2016,Krakowiak2018}. Importantly, Russo et al.~\cite{Russo2025softmodes} find that many biological control processes use a small number of linear combinations of states, of the form $C$, to initiate feedback mechanisms. In other words, they find that typically one or linear combinations of the sort in $C$ are used for feedback and control. Higher dimensional control, where many linear combinations like $C$ are used, may be selected out evolutionarily. Given this, in what follows, we focus our efforts mainly on cases where the number of linear combinations controlled, $D$, is either 1 or 2. Our thermodynamic findings from the previous sections lay out how stable feedback can be reliably achieved in many cases.

Finally, while our framework is most directly applicable to a system of non-equilibrium unimolecular (i.e., first-order or pseudo-first-order) reactions, we anticipate that it can be extended to certain classes of bimolecular reactions as well. In Refs.~  \citenum{gunawardena2012linear, nam2022linear, owen2023thermodynamic}, for example, it is described how under approximations like timescale separation bimolecular reaction networks or enzyme catalyzed reactions describing the dynamics of chemical concentrations can be cast into effective linear forms for which the matrix-tree theorem applies and our thermodynamic constraint holds.  These works also account for processes like degradation and synthesis within the reaction network (not just through the feedback rules), broadening the set of biological scenarios to which our results apply.  In truly non-linear reaction networks it is also sometimes possible to derive similar constraints on monotonic responses. \color{black}

We next show how systems with the local feedback rule in Equation \ref{eq1Ddyn} achieve control in a range of settings.


\begin{figure*}
\centering
\includegraphics[width=0.9\textwidth]{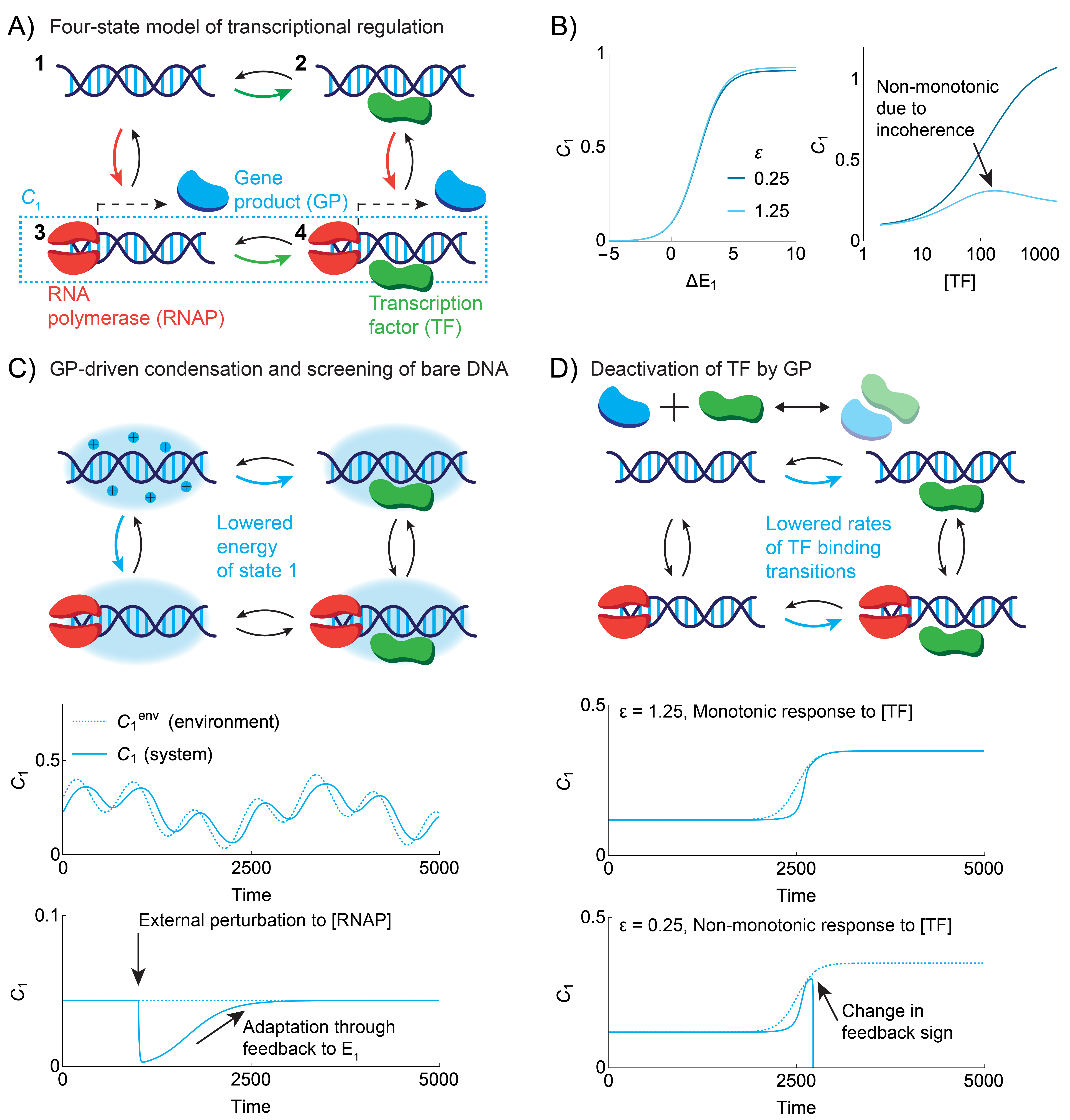}
\caption{
\color{\editcolor}
\textbf{Example of local biomolecular feedback control.}  
A) Schematic illustration of local feedback regulation of a DNA transcription site by the gene product of transcription. Throughout, we parameterize the base rates according to the values in Ref.~\citenum{martinez2025emergence}.  Feedback regulation enters as multiplicative modifications to these base rates as discussed Appendix \ref{App:NumMeth}. The observable is defined as $C_1 = p_3 + \epsilon p_4$.
B) Non-equilibrium response curves $C_1(\bp(\Delta E_1))$ and $C_1(\bp([\text{TF}]))$.
C) Condensate-mediated regulation of $E_1$, affecting rates $W_{21}$ and $W_{31}$.  Bottom plots show tracking of a time-varying environmental signal $C_1^\text{env}(t)$, and adaptation to a sudden perturbation $[\text{RNAP}] \rightarrow 0.05$.  The small plus symbols represent ions.  Throughout this figure $S_{a1} = 0.1$ and the remaining parameters are in Appendix \ref{App:NumMeth}.
D) GP-mediated regulation of TF and its effect on rates $W_{21}$ and $W_{43}$. Bottom plots show stable tracking ($\epsilon = 1.25$) and unstable tracking ($\epsilon = 0.25$) of an environmental signal $C_1^\text{env}(t)$. 
\color{black}
}
\label{BiologicalExamples}
\end{figure*}

\section{Tracking and adaptation with one regulated variable}
\color{\editcolor}
We first consider cases with one regulated observable $\lambda_a$, i.e., when $D=1$. This variable regulates through Equation~\ref{eq1Ddyn} a network observable $C_1$. The so-called multiplicity $M$ of $\lambda_a$ is set by the details of the biological problem. For concreteness, we first primarily study the four-state transcriptional circuit model in Figure \ref{BiologicalExamples}A, a common model explored for example in Refs.~\citenum{mahdavi2024flexibility, martinez2025emergence}, although we emphasize that rules of this type can work for larger and more disorganized chemical reaction networks.  In this model, transitions $1\rightarrow 2$ and $3\rightarrow 4$ have rates that linearly depend through pseudo-first-order kinetics on the ambient concentration of TF, and transitions $1\rightarrow 3$ and $2\rightarrow 4$ have rates that similarly depend on the concentration of RNAP.  The observable $C_1$ for this system is defined as 
\begin{equation}
\label{eq:c1def}
C_1=p_3 + \epsilon p_4
\end{equation}
defined as the weighted occupancy of the RNAP-bound states, and $\epsilon$ (playing the role of a $q_m$ parameter in Equation~\ref{eqCdefOrig}) reflects the relative change in transcription rate due to the bound TF.   

We consider two biophysical mechanisms by which the rates of the transcriptional circuit may be dynamically regulated. First, based on experimental studies of RNA-induced condensate formation around transcriptional loci \cite{henninger2021rna}, we model charge screening of the bare DNA through condensate formation as decreasing the energy $E_1$ of state 1.  From the constraints identified in the previous section, $C_1$ must vary monotonically with $E_1$ for any choice of $\epsilon$, as verified in Figure~\ref{BiologicalExamples}B. This corresponds to a $\lambda_a$ multiplicity of $M=1$.   Second, we treat regulation of the rates through changes in TF concentration. Because $[\text{TF}]$ affects two transitions through direct rate modulation (multiplicity $M=2$), monotonicity of $C_1(\bp([\text{TF}]))$ is not guaranteed and depends on the coherence conditions. Fixing other rate constants based on the parameters in Ref.~\citenum{martinez2025emergence}, the monotonicity of $C_1(\bp([\text{TF}]))$ depends on $\epsilon$ (Figure~\ref{BiologicalExamples}B). As discussed in Ref.~\citenum{martinez2025emergence} and Appendix \ref{App:Incoherence}, for the model parameters used here the regime $\epsilon < 1$, indicating a strong inhibitory effect of RNAP transcription by TF binding, corresponds to violation of the coherence conditions so that $C_1$ may depend non-monotonically on $[\text{TF}]$.
 
\textit{Feedback from condensate-mediated energy perturbation is stable}.  We next treat feedback control dynamics of the form of Equation~\ref{eq1Ddyn} through these two biophysical mechanisms. To illustrate the condensate-mediated regulation of $E_1$, we take the GP, in this case transcribed RNA molecules, to have the following dynamics:
\begin{equation}
    \dot{[\text{GP}]} = r C_1(\mathbf{p}) - \gamma \label{eqDNAE1}
\end{equation}
where $r$ is the rate of GP production by transcription and $\gamma$ is an export or degradation rate. We assume these latter processes are mediated by enzymes saturated by substrate, so the rates are independent of the instantaneous nuclear concentration $[\text{GP}]$. We further take the local charge density around the transcriptional loci to increase, and the energy correspondingly to decrease, in proportion $[\text{GP}]$, i.e., $E_1 = \phi [\text{GP}]$ with $\phi<0$; this simplifies the dependence of condensate formation on RNA concentration explored in Ref.~\citenum{henninger2021rna} and could be relaxed in future work. Identifying $S =r \phi$ and $C_1^\text{env} = \gamma/r$, Equation~\ref{eqDNAE1} maps onto
\begin{equation}
    \dot{E_1} = S(C_1(\mathbf{p}) - C_1^\text{env})
\end{equation}
which is of the local imperfect form in Equation \ref{eq1Ddyn}.  In Figure~\ref{BiologicalExamples}C we demonstrate that these feedback dynamics are capable of tracking a time-dependent signal, represented as an externally controlled temporal variation in the enzymatic degradation rate through $C_1^\text{env}(t)$.  
This type of feedback control can also solve adaptation tasks, in which $C_1^\text{env}$ is fixed but the underlying kinetic rates $\bt$ are shifted.  Studies such as Ref.~\cite{henninger2021rna} have identified RNA-mediated condensation as a potential route to buffer environmental perturbations through feedback control, but what biochemical properties are needed to explain this are not yet well-characterized.  We show that monotonicity of the dependence of $C([E_1])$ is sufficient to allow for stable adaptation.  As an example we model the environmental perturbation as a sudden decrease in $[\text{RNAP}]$, affecting the rates $1\rightarrow3$ and $2\rightarrow 4$.  Although this shift temporarily moves $C_1(\mathbf{p})$ away from the fixed point $C_1^\text{env}$, the condensate-mediated feedback through $E_1$ allows the system to recover under the new rates. We emphasize that such adaptation through local feedback connections can work in networks with arbitrary topologies, which contrasts with previous works that consider specially designed motifs or network substructures or require specific assumptions about the system's response \cite{lan2012energy, tu2018adaptation, barkai1997robustness, alon2019introduction, araujo2018topological, araujo2023universal, ma2009defining, aoki2019universal, PRXLife.3.013017, briat2016antithetic}. The success of these feedback dynamics requires that the environmentally specified value $C_1^\text{env}$ be reachable through variation of $\lambda$ within its allowed range. If $C_1$ still deviates from $C_1^\text{env}$ at saturating values of $\lambda$, the drift of $\lambda$ in Equation \ref{eq1Ddyn} will not terminate (until the underlying biophysical model assumptions break down). In other words, adaptation to a change in kinetic rates will not perfectly recover the original set point if, under the new rates, that set point is unreachable through variation in $\lambda$. We illustrate this in Appendix \ref{App:Adaptation}.

\textit{Stability through TF-mediated feedback depends on coherence}.  To illustrate the TF-mediated feedback dynamics, we model the GP as a molecule such as a translated protein product that binds to and deactivates the TF, so that only the complexed form TF:GP can bind to DNA. We take the binding reaction to be fast, so that the active concentration $[\text{TF}]$ is instantaneously given by
\begin{equation}
    [\text{TF}] = [\text{TF}]_\text{tot}\frac{K_\text{eq}}{K_\text{eq} + [\text{GP}]} \label{eqseq}
\end{equation}
where $[\text{TF}]_\text{tot}$ is the total TF concentration and $K_\text{eq}$ is the equilibrium binding constant.  Because this reaction is assumed fast, it does not contribute appreciable terms to the source-sink dynamics for GP in Equation \ref{eqDNAE1} as TF rapidly adjusts to the slowly varying GP concentration. As in Equation \ref{eqDNAE1}, we assume the dynamics of $[\text{GP}]$ are governed by production at rate $rC_1$ and degradation at rate $\gamma$. The dynamics for $[\text{TF}]$ are then
\begin{equation}
    \dot{[\text{TF}]} = S \frac{K_\text{eq}}{(K_\text{eq} + [\text{GP}])^2} (C_1(\mathbf{p}) - C_1^\text{env}) \label{eqTFfullfeedback}
\end{equation}
where $S = -r [\text{TF}]_\text{tot}$ and $C_1^\text{env} = \gamma/r$. These dynamics are of the form of Equation \ref{eq1Ddyn} for a $\lambda$-like variable, with a state-dependent prefactor arising from the nonlinear dependence of $[\text{TF}]$ on $[\text{GP}]$. In Figure \ref{BiologicalExamples}D we show that, when $\epsilon = 1.25$ (when the coherence conditions are satisfied), the response $C_1(\bp([\text{TF}]))$ is monotonic and these local feedback dynamics allow the system to track a time-varying $C_1^\text{env}(t)$. Convergence to the instantaneous $C_1^\text{env}(t)$ is not uniform along the trajectory due to the nonlinear prefactor, but the dynamics are stable. When $\epsilon = 0.25$ and $C_1(\bp([\text{TF}]))$ is non-monotonic, however, we observe divergence: the system enters a regime of the response curve where the dynamics run away due to positive feedback. We later show how such non-monotonic response can be stabilized by oppositely signed feedback signals on each reaction pathway affected by $[\text{TF}]$ (Figure~\ref{TFStabilization}). These results show that simple biophysical feedback connections from the occupancy of a Markov graph observable onto its various edge rate parameters can lead to stable tracking and adaptation to external signals.  Although the four-state DNA transcription model is topologically simple, similar results based on the monotonicity of non-equilibrium responses can also hold in arbitrarily complex networks (see Figures~\ref{2DLearning}, \ref{SI_TimescaleInstability}, and \ref{SI_Adaptation} for a more complex network example) including in transcriptional models with expanded state spaces \cite{bintu2005transcriptional, mahdavi2024flexibility}.

\textit{Feedback in more complex architectures}.
We emphasize that these results carry over to complex architectures beyond the model studied here. For example, transcription factor networks can function via other mechanisms that result in larger Markov networks~\cite{mahdavi2024flexibility}. As long as $D=1$ and coherence conditions are maintained, our results suggest that simple feedback rules of the sort in Equation~\ref{eq1Ddyn} are sufficient without any additional fine-tuning.   
\color{black}

\begin{figure*}
\centering
\includegraphics[width=\textwidth]{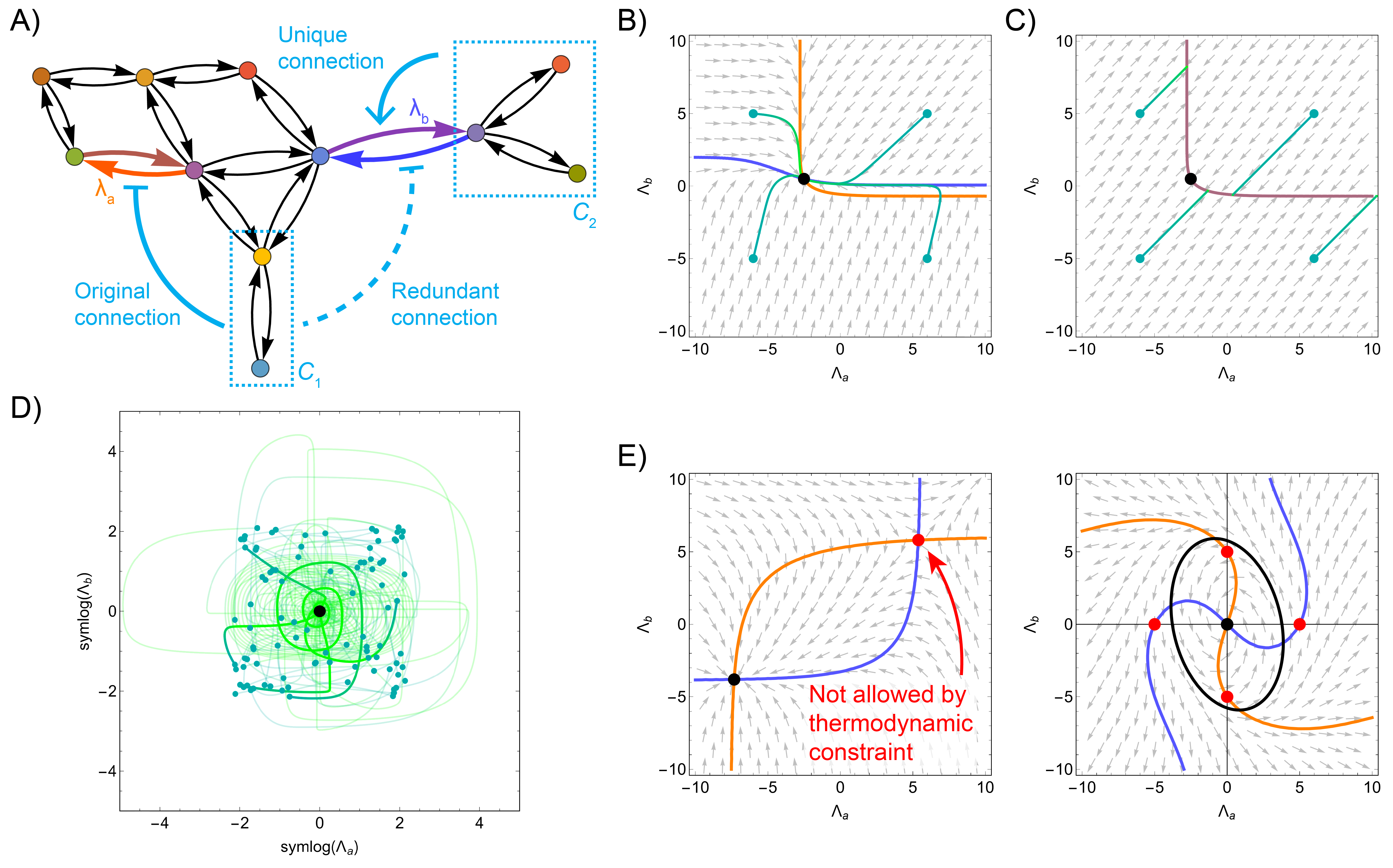}
\caption{\textbf{Stability of imperfect local feedback connections in two dimensions.} A) A random Markov network with two feedback connections; see Appendix \ref{App:NumMeth} for details on random graph generation. The original connection links observable $C_1$, defined as the average of the enclosed nodes, to $\lambda_a$ on edge $i\leftrightarrow j$. Two additional connections are shown: one (unique) links $C_2$, similarly defined, to $\lambda_b = \exp\Lambda_a$ on edge $k\leftrightarrow l$; and one (redundant, dashed) links $C_1$ to $\lambda_b = \exp \Lambda_b$. Regulated edges are colored orange and blue, with red-shaded arrows next to the labels indicating positive $F_{ij}$ contributions.
B) Phase plane for the original and unique connections under the adiabatic dynamics in the space $\Lambda_a = F_{ij}/2$, $\Lambda_b = F_{kl}/2$. Nullclines for $\dot{\Lambda}_a$ and $\dot{\Lambda}_b$ are drawn in orange and blue, respectively, and gray arrows indicate the direction of the flow field. Four trajectories (dark to light green in time, with initial conditions shown as points) are plotted, and the stable fixed point is shown in black.
C) As in panel B, but for the original and redundant connections. The nullclines for $\dot{\Lambda}_a$ and $\dot{\Lambda}_b$ overlap in this case.
 D) 100 trajectories from randomly sampled initial conditions on random graphs with $N_\text{n} = 11$ and $N_\text{e} = 13$. We plot on a symlog scale to show the full range of all trajectories, and we color trajectories with green at later times.  All trajectories converge to the stable fixed point at $\mathbf{0}$.
 E) Schematic illustration of how thermodynamic constraints on the monotonicity of observables along the $\Lambda_a$ and $\Lambda_b$ axes, combined with their unique intersections, preclude certain unstable structures such as limit cycles and saddle points. In these toy phase portraits, red points mark nullcline features that are forbidden by the thermodynamic constraints on our Markov network feedback control.
}
\label{2DLearning}
\end{figure*}

\section{Tracking and adaptation with two regulated variables}
We next consider the feasibility of control using local imperfect feedback on two regulated observables, tuned using two rate parameters $\Lambda_a$ and $\Lambda_b$. Introducing control over an additional driving force presents a challenge, because now monotonicity is not guaranteed for any type of rate perturbations as the sign of the derivative $\partial \pi_k / \partial \Lambda_a$ in general depends on the value of $\Lambda_b$. Despite this, we provide numerical evidence and analytical arguments demonstrating that local stability also implies global stability for the local imperfect feedback dynamics in two dimensions.

Figure \ref{2DLearning}A shows a randomly drawn Markov network with two regulated edges.  Here we take the rate perturbations to be on the affinity-like parameters $F_{ij}$ on these edges; see Appendix \ref{App:SpecificPerturbations} for a biophysical example of such antisymmetric rate modulations.  Feedback on these two edge affinities can be implemented in two general ways: (a) using two different observables, $C_1$ and $C_2$, i.e.,
\begin{eqnarray}
    \dot{\Lambda}_a &=& S_{a1}\left(C_1- C_1^\text{env}\right) \label{eq2d1} \\
    \dot{\Lambda}_b &=& S_{b2}\left(C_2- C_2^\text{env}\right), \label{eq2d2}
\end{eqnarray}
or (b) redundantly using the same observable, i.e.,
\begin{eqnarray}
    \dot{\Lambda}_a &=& S_{a1}\left(C_1- C_1^\text{env}\right) \label{eqred1} \\
    \dot{\Lambda}_b &=& S_{b1}\left(C_1- C_1^\text{env}\right). \label{eqred2}
\end{eqnarray}
These two approaches give rise to qualitatively different feedback dynamics. When two distinct observables are used, the system typically possesses a unique fixed point, located at the intersection of the contours $C_1(\Lambda_a, \Lambda_b) = C_1^{\text{env}}$ and $C_2(\Lambda_a, \Lambda_b) = C_2^{\text{env}}$, which coincide with the nullclines $\dot{\Lambda}_a = 0$ and $\dot{\Lambda}_b = 0$. In Appendix \ref{App:UniqueFixed} we demonstrate that any isolated fixed point is unique for all types of rate perturbations, meaning there are no other roots of the equations $\dot{\Lambda}_a = 0$ and $\dot{\Lambda}_b = 0$. By appropriately selecting the values of $S_{a1}$ and $S_{b2}$, the fixed point can be made locally stable. Figure~\ref{2DLearning}B illustrates this local stability through several convergent trajectories and a vector field indicating the direction of local flow.

\begin{figure*}[ht!]
\begin{center}
\includegraphics[width=\textwidth]{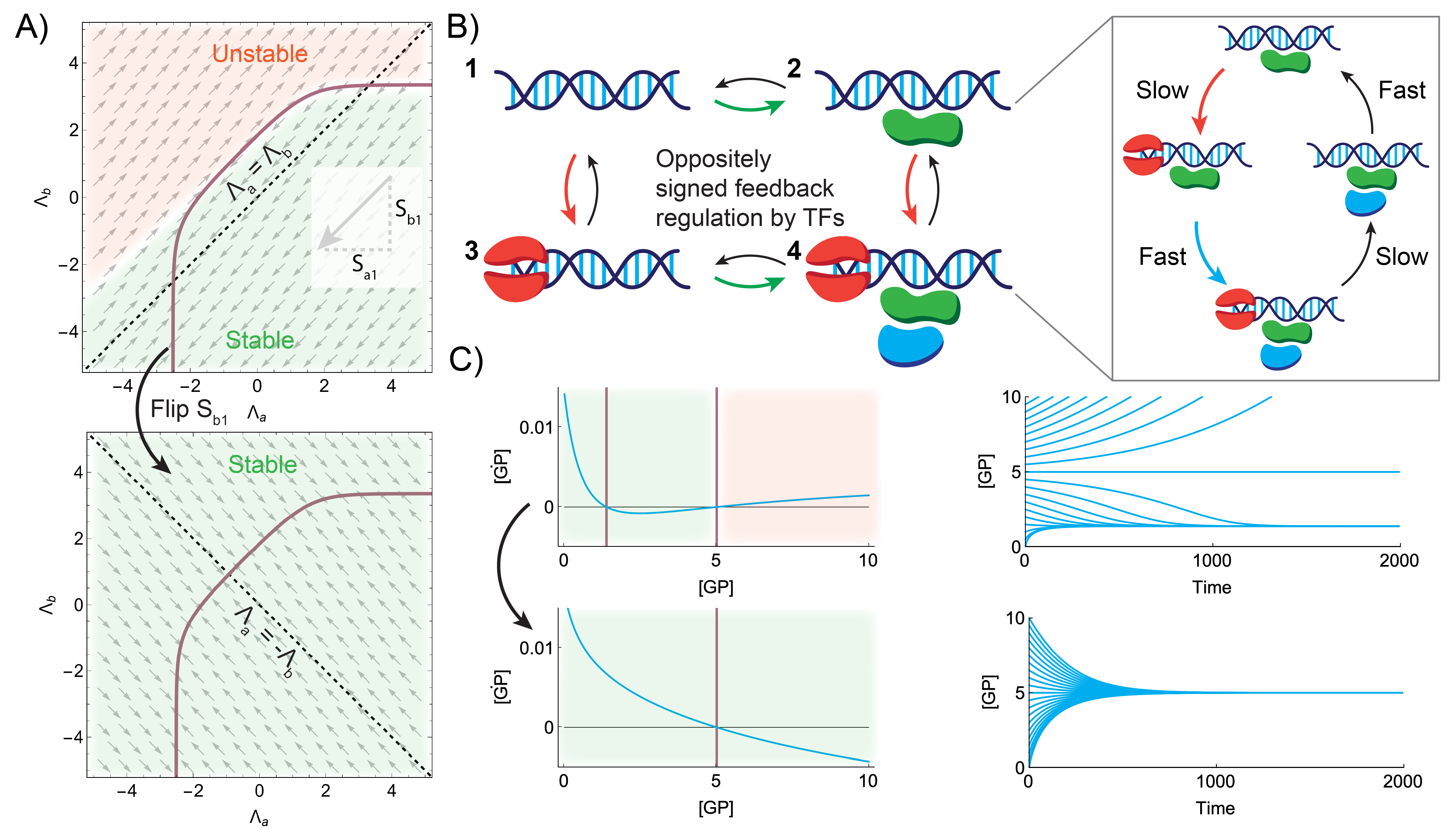}
\caption{
\color{\editcolor}
\textbf{Stabilizing non-monotonic responses.}  
A) Flow fields for the adiabatic feedback dynamics in a system where one variable affects two transition rates $(D=1,\  M=2)$. The top plot shows the case in which feedback on this variable produces unstable regions that do not globally flow toward the purple nullcline. The bottom plot shows how flipping the feedback sign on the second appearance of this regulatory variable can globally stabilize the feedback dynamics.
B) The same system as in Figure \ref{BiologicalExamples}D, but modified to show how the second reaction pathway regulated by TF may take feedback of the opposite sign from the GP. The inset shows how one may expand the transitions $2 \leftrightarrow 4$ to effectively recover the original kinetic rates through rate-limited two-step reaction pathways.
C) \textit{Top left}: Phase plot of $\dot{[\text{GP}]}$ versus $[\text{GP}]$ in the original kinetic scheme, illustrating regions in which the flow is unstable due to a non-monotonic response $C_1(\bp([\text{TF}]))$ with $\epsilon = 0$.
\textit{Top right}: Trajectories initialized at different values of $[\text{GP}]$, illustrating the lack of global stability.
\textit{Bottom row}: Same as the top row, but for the altered kinetic scheme in panel B, illustrating the recovery of global stability through the altered feedback dynamics on the non-monotonic variable $[\text{TF}]$.
\color{black}
}
\label{TFStabilization}
\end{center}
\end{figure*}

If, instead, the same observable $C_1$ is redundantly used as feedback for both $\Lambda_a$ and $\Lambda_b$, the system possesses a one-dimensional manifold of fixed points defined by the solution space of $C_1(\Lambda_a, \Lambda_b) = C_1^{\text{env}}$. This redundant feedback could correspond in a biological example to a single DNA transcription region encoding gene products which modulate two transcription factor activities $\Lambda_a$ and $\Lambda_b$ in a shared regulatory network.  With appropriately chosen $S_{a1}$ and $S_{b1}$ the trajectories converge stably toward the one-dimensional manifold of fixed points (Figure \ref{2DLearning}C). This redundancy can be biologically advantageous by simplifying the flow fields and enabling more direct convergence to $C_1 = C_1^\text{env}$, minimizing large excursions in $\bL$ space. Although one-dimensional solution manifolds are always a feature of redundant feedback connections, we show in Appendix \ref{App:UniqueFixed} that they can also emerge for certain ``low-rank'' choices of two distinct observables, $C_1$ and $C_2$.  We note that, in addition to tracking tasks, adaptation is also possible in the case of two-dimensional control, but for brevity we do not show this here.

Figure \ref{2DLearning}D illustrates a numerical search over random graphs and random initial conditions around a locally stable fixed point. In every case, the initial conditions converge to the fixed point, suggesting that the basins of attraction for these fixed points extend throughout the entire domain sampled, implying that the fixed points are globally stable.  While rigorously proving global stability in two dimensions is challenging, we argue that it is plausible due to the fact that the nullclines are significantly limited by the non-equilibrium thermodynamic constraints.  Specifically, the uniqueness of the isolated fixed points ensures that no two nullclines can intersect more than once (see Appendix \ref{App:UniqueFixed} for a proof of unique intersections), preventing the formation of saddle points with unstable manifolds (Figure \ref{2DLearning}E) \cite{perko2013differential, izhikevich2007dynamical}.  Furthermore, the constraint on the monotonicity of the observables ensures that the nullclines intersect any vertical or horizontal line in $\bL$ space at most once. This precludes the formation of unstable limit cycles around the stable fixed point.  These restrictions on the formation of unstable structures suggest that the two-dimensional dynamics, which in general are significantly more constrained than three-dimensional systems \cite{ott2002chaos}, are globally stable as observed numerically. Thus, although simultaneously modifying two driving forces can result in incorrect sign estimates $S_{a1}$ and $S_{b1}$ in certain regions of $\bL$, the topologically simple structure of the nullclines, enforced by the non-equilibrium thermodynamic constraint, appears to ultimately guarantee global stability of the local feedback dynamics.

\color{\editcolor}
\textit{Stabilizing non-monotonic responses through feedback signs.}
The redundant feedback case (where one observable regulates two edge variables) is similar to a single variable affecting two edges simultaneously, as in the TF-regulated edges with $M=2$ in Figure \ref{BiologicalExamples}A. The key difference is that for redundant connections to two independent variables (Equations~\ref{eqred1} and \ref{eqred2}), the prefactors $S_{a1}$ and $S_{b1}$ can be set independently, whereas if both connections represent a single physical variable these prefactors are not independent. In Figure \ref{TFStabilization}A we show the flow field for a single variable $\Lambda_a = \Lambda_b$ affecting two edges, visualized in the expanded space where these are treated as independent. Along the constraint line, a non-monotonic response produces an unstable region, leading to divergent trajectories as in Figure~\ref{BiologicalExamples}D. If possible, changing the sign of $S_{b1}$ would rotate the flow field to point toward the nullcline along the new constraint line $\Lambda_a = -\Lambda_b$, which suggests a biological strategy for stabilizing non-monotonicities by modulating how each affected edge responds to the regulated variable.

To illustrate this, we return to the example of Figure \ref{BiologicalExamples}D, in which a TF regulates two edges of a four-state transcriptional circuit, now modified as in Figure \ref{TFStabilization}B. At $\epsilon = 0$ the response $C_1(\bp([\text{TF}]))$ is non-monotonic, and the adiabatic feedback dynamics, in which active TF concentration depends on GP through Equation \ref{eqseq}, contain unstable regions (Figure \ref{TFStabilization}B). We now modify the kinetic model so that state 4 is bound by the TF:GP complex rather than free TF (as in state 2).  We model the transitions $2\rightarrow 4$ and $4\rightarrow 2$ as effective two-step reactions which are rate-limited in a manner that recovers the original transition rates; see the inset of Figure \ref{TFStabilization}B. This effectively flips the sign of feedback regulation on transition $3\rightarrow 4$, in the spirit of the stabilization strategy of Figure \ref{TFStabilization}A. As shown in Figure \ref{TFStabilization}C, the feedback dynamics through GP are now globally stable. Thus, non-monotonicities arising from a single variable appearing on multiple edges can be stabilized by modifying the kinetic scheme so that the two edge regulations respond with opposite sign.
\color{black}

\section{Discussion}
We have studied the problem of controlling non-equilibrium  processes, modeled generically as Markov jump processes. Such models are routinely used to describe a variety of non-equilibrium biological and synthetic processes. Our results reveal a simple, broadly applicable strategy for achieving tracking and adaptation in non-equilibrium biophysical and synthetic systems through biologically plausible feedback mechanisms.  The key to control and adaptation is a thermodynamic constraint that ensures that locally stable feedback connections remain globally stable for up to two regulated driving forces and imposes structural limitations on the global flow field in higher-dimensional systems. While these imperfect local feedback mechanisms do not yield exact solutions to the problems of control and adaptation, they are surprisingly effective across a wide range of conditions and network architectures.

This control framework could be relevant to a range of biophysical processes.  The decision made by viruses to undergo either lysis or lysogeny at a collective level has been shown to involve a feedback-regulated maintenance of arbitrium peptides in the surrounding medium \cite{erez2017communication}. Such switch-like can behavior could be obtained using Equation~\ref{eq1Ddyn} with $C_1^{\text{env}}$ encoding a sigmoidal dependence on arbitrium levels.  The robustness of biochemical oscillations to temperature variations and phase shifts in an entraining signal may also involve dynamically stable feedback connections between the environment and edge rates \cite{fei2018design, fu2024temperature}.  
\color{\editcolor}
Additionally, the observed robustness of heat-shock response efficacy to model specification may be partly explained by a thermodynamically constrained response, which would allow several distinct models to exhibit adaptive behavior \cite{zheng2016dynamic, krakowiak2018hsf1, garde2023transcriptional}.
\color{black}
Finally, adaptive responses of the immune system are thought to near-optimally track the evolution of a diverse range of antigens \cite{mayer2019well, kepler1993somatic, nourmohammad2021optimal}. Our work could help show how minimal feedback loops which are based on locally estimated antigen binding affinities, ensure that an averaged readout over the antibody sequences robustly reflect the presented antigens. Another key insight from our analysis is the connection between our findings and the general dynamical properties of biochemical adaptation. Specifically, our results are consistent with previous studies on adaptation mechanisms that depend on constraints imposed on the partial derivatives of system variables \cite{lan2012energy, tu2018adaptation} or on network topology \cite{araujo2018topological, araujo2023universal, ma2009defining, aoki2019universal, PRXLife.3.013017, briat2016antithetic}. These constraints, which must be assumed to hold in biochemical models of adaptation, establish relationships between response functions and network structure that facilitate integral feedback, enabling perfect adaptation. By contrast, we derive a thermodynamic constraint which, without assuming any other network properties, enforces monotonicity in non-equilibrium responses and allows integral feedback loops to achieve stable regulation without requiring precise parameter tuning.  

The chemical reaction networks considered here are linear models. While our results may extend to certain classes of bimolecular reaction networks \cite{gunawardena2012linear, nam2022linear, owen2023thermodynamic}, it remains an open question whether analogous thermodynamic constraints can be formulated and exploited in fully nonlinear chemical dynamics. Nonlinearities in reaction dynamics can introduce non-monotonic responses \cite{floyd2025limits}, yet we anticipate that broad classes and parameter regimes of nonlinear networks will still exhibit monotonic responses that can be leveraged to achieve dynamically stable feedback without reliance on specific architectural features. As we plan to explore more in future work, one can also establish monotonicity constraints in certain genuinely non-linear kinetic models, hence extending the range of these feedback dynamics to a class of bimolecular reaction networks.   At the same time, non-monotonic responses may themselves be advantageous, for example in constructing bistable feedback dynamics relevant to biological decision-making. Such non-monotonic behavior can arise either from nonlinear reaction dynamics or, as shown in Refs.\ \citenum{mahdavi2024flexibility, floyd2025limits}, from linear systems in which the input variable modulates network rates through multiple pathways.

\color{\editcolor}

The specific source-sink feedback dynamics for the regulated variable 
(Equation~\ref{eq1Ddyn}) analyzed in the main text are illustrative of settings in which thermodynamic constraints can ensure stability, rather than exhaustive. More generally, quantities such as $C_1 - C_1^\text{env}$ can serve as feedback components within broader and more complex feedback architectures, and the guarantees on the signf of the relevant partial derivatives can help to ensure stability in many of these settings. For example, although the functional form in Equation~\ref{eqTFfullfeedback} differs from that in Equation~\ref{eq1Ddyn} by a nonlinear prefactor, this prefactor contributes only a monotonic scaling that does not affect stability. More generally, one may target rates such as $C_1 - \gamma \Lambda_a$, with $\gamma$ a first-order reaction rate, through analogous feedback rules. Future work could explore how these core results extend to a broader class of biologically realistic control strategies.

Our analytical results establish global stability of the local imperfect feedback dynamics for $D=1$ and provide numerical evidence and analytical arguments for $D=2$, but global stability is not guaranteed for $D \geq 3$, consistent with the absence of a Poincar\'{e}--Bendixson theorem in higher dimensions \cite{perko2013differential, ott2002chaos}. In three dimensions and above, the topological constraints that enforce convergent behavior in 2D are absent, permitting a richer class of dynamics for which stability guarantees do not hold in general. The biologically relevant regime may in fact lie within the low-dimensional domain where our results apply. Russo et al.~\cite{Russo2025softmodes, Russo2025review} show that many well-characterized homeostatic systems employ controllers that integrate diverse signals into one or two molecular concentrations before taking action, with examples including (p)ppGpp in bacteria~\cite{Traxler2008}, cAMP-PKA in yeast~\cite{Zaman2008, Conrad2014}, and Hsf1~\cite{Zheng2016, Krakowiak2018}, all of which operate in regimes with $D=1$. The authors further show that evolutionary selection for homeostasis favors such low-dimensional controllers over more complex alternatives, suggesting that even in networks with many regulated edges, the number of independent control variables $D$ is small, placing the system in the regime where the thermodynamic constraints derived here may guarantee stability. Characterizing the input multiplicity $M$ and the coherence properties of these biological controllers within a Markov network description, as well as understanding which topological features are important for reachability and global stability in arbitrary dimension, remain important directions for future work.

Can add connection to anti-kinetic proofreading work

\color{black}
A number of recent works have established thermodynamic constraints on the responses of Markov networks, drawing on tools such as the Markov tree theorem to relate steady-state observables to network structure~\cite{liang2024thermodynamic, aslyamov2024general, aslyamov2024nonequilibrium, aslyamov2025nonequilibrium, qureshi2026thermodynamic, harunari2024mutual, owen2020universal, owen2023thermodynamic, owen2023size, arunachalam2025information, bebon2026mutual}. A common theme in these works is bounding the magnitude or sensitivity of responses, for instance by establishing inequalities on fluxes or dissipation rates under external perturbation. Our focus is complementary but distinct: we are primarily concerned with determining the sign of steady-state responses, including in cases where multiple variables are regulated simultaneously, and with leveraging these constraints for the purpose of control. Other control-related results in the literature are derived in an open-loop setting and make use of counterdiabatic driving techniques, in which driving protocols are prescribed externally~\cite{ilker2022shortcuts, iram2021controlling, patra2017shortcuts, rahav2008directed}. Our work instead studies closed-loop feedback control, in which a biological or engineered system actively regulates edge parameters in response to measured outputs. This distinction qualitatively alters the corresponding non-adiabatic perturbation theory (Appendix~\ref{App:Non-adia}) and places our results closer to the literature on biochemical adaptation.
\color{black}

Our results can also be compared to the usefulness of imperfect gradients in machine learning, where optimization processes often succeed despite using gradient estimates that only loosely approximate the true global cost function. Techniques such as semi-gradients in temporal difference reinforcement learning (RL) and stochastic mini-batch gradients rely on the fact that even a rough alignment with the true gradient is sufficient for descent \cite{lecun2015deep, sutton2018reinforcement}.   We note that stochastic gradient estimates typically have no systematic errors, whereas other gradient approximations and our imperfect gradients would be expected to.  Similarly, methods like contrastive divergence train generative models by estimating equilibrium gradients using samples from partially converged distributions \cite{carreira2005contrastive, decelle2021equilibrium, agoritsas2023explaining}. Even in feed-forward neural networks, backpropagation can tolerate severe approximations; remarkably, networks can still learn effectively when random backward weights replace the true error derivatives, provided that forward passes remain accurate \cite{lillicrap2016random, lillicrap2020backpropagation, wright2022deep}. Our work extends these ideas to biophysical systems by showing that in such systems, highly localized and imprecise feedback can still enforce structured, stable dynamics due to fundamental thermodynamic constraints which preserve an overlap with a universal globally stable dynamics. 

Finally, our results have broader implications for both synthetic biology and evolutionary theory. The effectiveness of imperfect local feedback raises the possibility that ancestral biological systems may have relied on similarly approximate control mechanisms before refining them through evolutionary processes. This aligns with the idea that evolvability favors simple, modular control structures that are easy to modify and recombine. From a synthetic biology perspective, our findings suggest that designing robust feedback regulation may not require precise circuit engineering, but instead may leverage the inherent constraints of non-equilibrium dynamics to achieve stability and adaptability. This opens the possibility of engineering synthetic biochemical systems with minimal design complexity while still ensuring reliable function; see Refs. \citenum{chen2024synthetic, yang2025engineering} for recent examples. Understanding how thermodynamic constraints shape the feasibility of control in biochemical systems could thus provide guiding principles for both studying evolutionary processes and designing biomimetic regulatory networks.

\appendix 

\color{\editcolor}
\section{Proof of the coherence conditions}\label{App:CoherenceProofs}
Here we establish that the coherence conditions, Equations \ref{eqmonorderinc} and \ref{eqmonorderdec}, are sufficient to ensure monotonicity of the function $\pi_m(\lambda)$, Equation \ref{eqpimlambda}.  We first differentiate 
\begin{equation}
\frac{\partial \pi_m}{\partial \lambda} = \sum_\mu \theta_\mu^m\,\frac{\partial w_\mu}{\partial\lambda}. \label{eqpiderivlambda}
\end{equation}
Defining $Z(\lambda)\equiv\sum_{\nu} \bar{\zeta}_\nu \lambda^\nu$, the weight derivatives are
\begin{equation}
\frac{\partial w_\mu}{\partial \lambda}
=
\frac{
\mu \bar \zeta_\mu \lambda^{\mu-1} Z
-
\bar \zeta_\mu \lambda^\mu Z'
}{Z^2}
\end{equation}
where $Z' = \sum_\nu \bar{\zeta}_\nu \nu \lambda^{\nu-1}$.
Substituting, we have
\begin{equation}
\frac{\partial \pi_m}{\partial\lambda}=
\frac{1}{Z^2}
\sum_{\mu,\nu}
\bar \zeta_\mu \bar \zeta_\nu
\lambda^{\mu+\nu-1}
(\mu-\nu)\theta_\mu^m .
\end{equation}
This expression is anti-symmetric under exchange of the indices $\mu$ and $\nu$, so we form the symmetric average 
\begin{align}
\frac{\partial \pi_m}{\partial\lambda}
&=
\frac{1}{2Z^2}
\sum_{\mu,\nu}
\bar \zeta_\mu \bar \zeta_\nu
\lambda^{\mu+\nu-1}
(\mu-\nu)
\left(
\theta_\mu^m-\theta_\nu^m
\right).
\end{align}
We can now restrict this symmetric double sum to run over the ordered pairs $\mu<\nu$
\begin{equation}
\frac{\partial \pi_m}{\partial\lambda}
=
\frac{1}{Z^2}
\sum_{\mu<\nu}
\bar \zeta_\mu \bar \zeta_\nu
\lambda^{\mu+\nu-1}
(\nu-\mu)
\left(
\theta_\nu^m-\theta_\mu^m
\right).
\end{equation}
From this representation, it is clear that since $\nu - \mu > 0$ for each term, the sign for that term depends on the relative ordering of the $\theta_\nu^m$ and $\theta_\mu^m$ ratios.  The orderings of Equations \ref{eqmonorderinc} or \ref{eqmonorderdec} are therefore sufficient to ensure monotonicity.  

Because $\sum_\mu w_\mu(\lambda) = 1$ for all $\lambda$, we can from Equation \ref{eqpiderivlambda} write
\begin{equation}
    \frac{\partial \pi_m}{\partial \lambda} = \sum_\mu (\theta_\mu^m - L_m)\,\frac{\partial w_\mu}{\partial\lambda} \label{eqderivoffset}
\end{equation}
where $L_m$ is an arbitrary constant that does not depend on $\lambda$.  This will allow us to derive other facts about the non-equilibrium response.  For example, if $N_\lambda = 1$ we may set $L_m = \theta_0^m$ to obtain
\begin{equation}
    \frac{\partial \pi_m}{\partial \lambda} = (\theta_1^m - \theta_0^m)\,\frac{\partial w_1}{\partial\lambda}
\end{equation}
which implies the response for all nodes $m$ are proportional to each other, containing a node-specific prefactor that multiplies a shared component carrying the dependence on $\lambda$; for example, Equation \ref{eqRmm} holds for such responses.

\section{Monotonicity and proportionality of various perturbations}\label{App:SpecificPerturbations}
The framework described in Section \ref{sec:ThermoConstraints} is for a generic perturbation to the network rates parameterized by some variable $\lambda$.  Here we consider several specific types of perturbation and establish monotonicity of their non-equilibrium responses.  We also show the non-equilibrium responses at the different nodes $m$ are in many cases proportional to each other, containing a node-specific prefactor which multiplies a shared part depending on $\lambda$.  See the text after Equation \ref{eqmarkovrate} for definitions of how the parameters $X_{ij}$, $E_j$, $B_{ij}$, and $F_{ij}$ affect the rates $W_{ij}$ and $W_{ji}$.

\subsection{Perturbing edge prefactors}
We first consider parameterizing an edge rate through $\lambda = \exp X_{ij}$.  From the matrix-tree expression we obtain an expression for the steady-state occupancy at node $m$
\begin{equation}
    \pi_m(\lambda) = \frac{\zeta^m_1 \lambda + \zeta^m_0}{\bar{\zeta}_1\lambda + \bar{\zeta}_0}. \label{eqsimplelambda}
\end{equation}
We thus have the case $N_\lambda = 1$, so the non-equilibrium response to this perturbation is monotonic and proportional for all nodes $m$.

An example of a biochemical process which may modulate an $X_{ij}$ term is the chemostatted concentration of a binding partner in a pseudo-first order chemical reaction.  The transcription factor concentrations in our analysis of the transcription regulatory network play a role like this.  

\subsection{Perturbing node energies}
We next consider perturbing the energy $E_j = \ln\lambda$.  The matrix-tree expression for this case yields the same form as Equation \ref{eqsimplelambda}, with $\zeta^j_1 = 0$ and $\zeta^m_0 = 0$ for all $m$.  This is because no trees flowing into node $j$ contain the factor $\lambda$ while all trees flowing into the other nodes do.  We again have the case $N_\lambda = 1$, so the non-equilibrium response to this perturbation is monotonic and proportional for all nodes $m$.

An example of a biophysical process which may modulate an $E_{j}$ term is the pH or ion-dependent stabilization of certain molecular configurations, which can shift the free energy level of an individual state and affect all rates out of it.

\begin{figure*}[ht!]
\begin{center}
\includegraphics[width=0.9\textwidth]{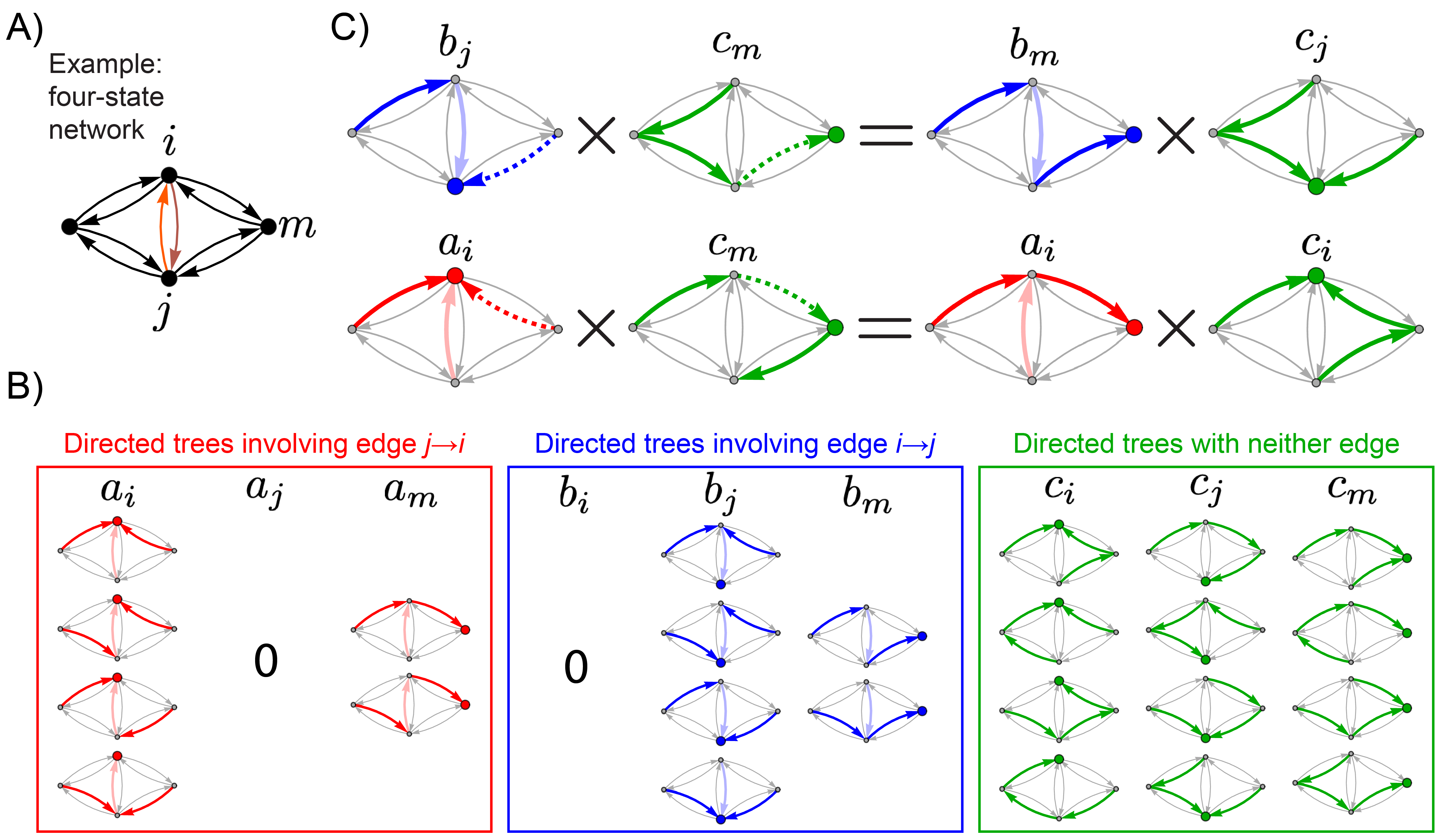}
\caption{\textbf{Illustration of the tree swapping procedure.}  A)  Example four-state network in which the rates $W_{ij}$ and $W_{ji}$, shown as orange arrows, are perturbed.  B)  Enumeration of the trees which contribute to the terms in Equations \ref{eqaibj}-\ref{eqmainkin}.   Faint red and blue arrows represent the edges $W_{ij}$ and $W_{ji}$ which have been factored out of the numerical tree weights.   C)  Illustration in a four-state graph of the correspondence between terms on both sides of Equation \ref{eqmainkin}.  Two schematic equations are depicted.  On the top, it is shown how a term from the sum $b_jc_m$ can be mapped, through swapping of the dashed edges, into a term from $b_m c_j$, while on the bottom it is shown how a term from the sum $a_ic_m$ can be mapped into a term from $a_m c_i$.}
\label{SI_TreeSwapping}
\end{center}
\end{figure*}

\subsection{Perturbing edge barriers}
We next consider perturbing an edge barrier $B_{ij}=-\ln\lambda$.  This case is more complex than those considered previously because two edges, $j\rightarrow i$ and $i\rightarrow j$, vary simultaneously as $B_{ij}$ changes.  The steady-state occupancy at node $m$, with the dependencies on edges $j\rightarrow i$ and $i\rightarrow j$ factored out, can be written as
\begin{equation}
\pi_m =\frac{a_m W_{ij} + b_m W_{ji} + c_m}{\bar{a} W_{ij} + \bar{b} W_{ji} + \bar{c}} \label{eqpikWij}
\end{equation}
where
\begin{align}
   a_m = \sum_{T_m\in \mathcal{Q}_{j\rightarrow i}} w^{-}(T_m) \geq 0
\end{align}
is a sum over all directed spanning trees $T_m$ rooted at node $m$ in the set $\mathcal{Q}_{j\rightarrow i}$ of trees containing the directed edge $j\rightarrow i$ (see Figure \ref{SI_TreeSwapping} for an example).  The quantity $w^{-}(T_m)$ is the product of all edge rates in this tree except for $W_{ij}$, which has been factored out (represented by the $^-$ superscript).  Similarly, we have
\begin{equation}
   b_m = \sum_{T_m\in \mathcal{Q}_{i\rightarrow j}} w^{-}(T_m) \geq 0
\end{equation}
for the directed spanning trees rooted at node $m$ in the set $\mathcal{Q}_{i\rightarrow j}$ of trees containing $i\rightarrow j$.  Note that the directed spanning trees which include $i\rightarrow j$ necessarily exclude $j\rightarrow i$, so the sums in $a_m$ and $b_m$ are over separate directed trees.  Finally, the sum over directed trees in which neither $i\rightarrow j$ nor $j\rightarrow i$ appear is 
\begin{equation}
   c_m = \sum_{T_m\in \mathcal{Q}_{(-)}} w(T_m)\geq 0
\end{equation}
where the sum is over directed trees in the set $\mathcal{Q}_{(-)}$ without $i\rightarrow j$ or $j\rightarrow i$, and whose path weights $w(T_m)$ have no terms factored out.  The coefficients in the denominator of Equation \ref{eqpikWij} include sums over all nodes $m'$:
\begin{align}
   \bar{a} =& \sum_{m'} a_{m'} > a_m, 
\end{align}
with $\bar{b}$ and $\bar{c}$ defined similarly.  We emphasize in defining these quantities that the node labels $i$ and $j$ refer to the neighbors of the regulated edge and $m$ refers to the observed node.  Thus, $a_m$, $b_m$, and $c_m$ depend on the choice of $i$ and $j$.  To keep notation minimal and since we will only consider one regulated edge unless specified otherwise, we do not adorn $a_m$ with the labels $i$ or $j$.  

Writing $W_{ij} = W_{ij}^0 e^{-B_{ij}}$ and $W_{ji} = W_{ji}^0e^{-B_{ij}}$, we have
\begin{equation}
    \pi_m =\frac{(a_m W_{ij}^0 + b_m W_{ji}^0)\lambda + c_m}{(\bar{a} W_{ij}^0 + \bar{b} W_{ji}^0)\lambda + \bar{c}}.
\end{equation} 
This expression is again of the form in Equation \ref{eqsimplelambda}, so the non-equilibrium response is monotonic and proportional for all nodes $m$.  

An example of a biochemical process which may modulate an $B_{ij}$ term is fast enzyme-catalyzed reaction of the form $i + E \leftrightarrow j + E$, with the concentration of the enzyme $E$ symmetrically scaling both transition directions through pseudo-first-order kinetic effects.

\subsection{Perturbing edge affinities}
Finally, we consider perturbing an edge affinity $F_{ij} = 2\ln \lambda$, which anti-symmetrically affects the logarithms of the rates $W_{ij}$ and $W_{ji}$.  Writing $W_{ij} = W_{ij}^0 e^{F_{ij}/2}$ and $W_{ji} = W_{ji}^0e^{-F_{ij}/2}$, from Equation \ref{eqpikWij} we have
\begin{eqnarray}
    \pi_m &=& \frac{a_m W_{ij}^0\lambda + b_m W_{ji}^0\lambda^{-1} + c_m}{\bar{a} W_{ij}^0\lambda + \bar{b} W_{ji}^0\lambda^{-1} + \bar{c}} \nonumber \\
     &=& \frac{a_m W_{ij}^0\lambda^2 + c_m \lambda +b_m W_{ji}^0}{\bar{a} W_{ij}^0\lambda^2 + \bar{c}\lambda + \bar{b} W_{ji}^0 }. \label{eqpimFijpert}
\end{eqnarray}
For this case we thus have $N_\lambda = 2$.  We will find that, surprisingly, the coherence conditions (Equation \ref{eqmonorderinc} or \ref{eqmonorderdec}) hold automatically for this perturbation due to a non-trivial relationship satisfied by the coefficients in Equation \ref{eqpimFijpert}.  

\begin{figure}[!ht]
\begin{center}
\includegraphics[width=\columnwidth]{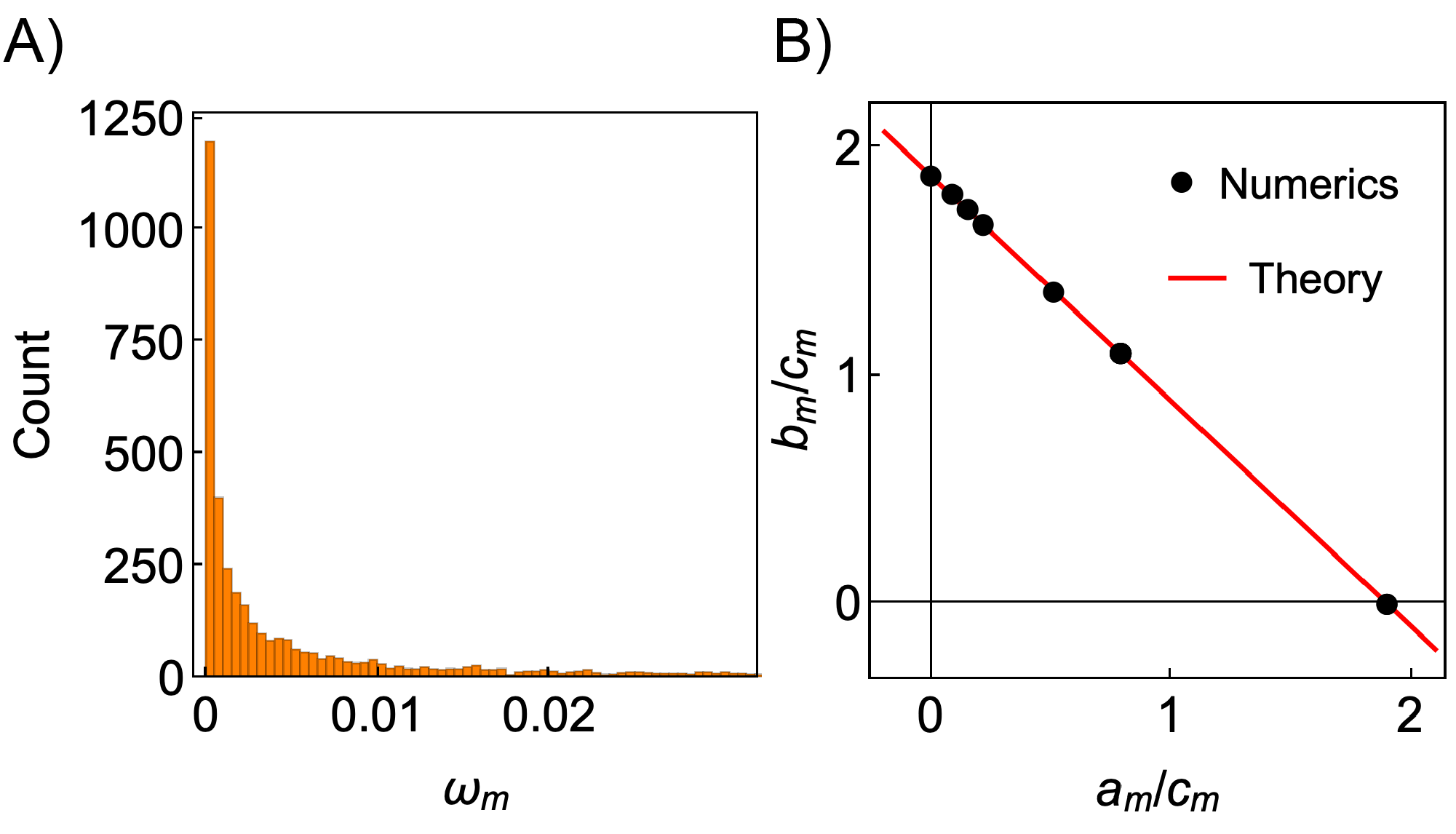}
\caption{\textbf{Numerical verification of tree weight constraints.}  A)  Histogram of $\omega_m \equiv (a_m/\bar{a} - c_m/\bar{c})(c_m/\bar{c} - b_m/\bar{b})$, which will be positive the if quantities in Equations \ref{eqxac} - \ref{eqxab} are either all positive or all negative,  for each node $m$ of 50 randomly generated graphs.  B)  Numerical verification of the theoretical result $b_m/c_m = b_j/c_j - (b_j c_i/a_i c_j) (a_m/c_m)$, which is implied by Equation \ref{eqmainkin}, for randomly sampled graph.  We compute the offset $b_j/c_j$ and slope $(b_j c_i/a_i c_j)$ from the rate parameters and plot this predicted line against computed values of $a_m/c_m$ and $b_m/c_m$.}
\label{SI_MonNumerics}
\end{center}
\end{figure}

We first observe that, because $\theta^m_2 = a_m / \bar{a}$, $\theta^m_1 = c_m/\bar{c}$, and $\theta^m_0 = b_m/\bar{b}$, the coherence conditions for Equation \ref{eqpimFijpert} are equivalent to the following quantities sharing the same sign:
\begin{eqnarray}
   x^{ac}_m &=& a_m \bar{c} - c_m \bar{a}, \label{eqxac} \\ 
   x^{cb}_m &=& c_m \bar{b} - b_m \bar{c}, \label{eqxcb} \\ 
   x^{ab}_m &=& a_m \bar{b} - b_m \bar{a}. \label{eqxab}
\end{eqnarray}
The shared sign of these terms follows from one of our key results (illustrated in Figure \ref{MonotonicityProof}):
\begin{equation}
    a_i = b_j, \label{eqaibj}
\end{equation}
\begin{equation}
    a_j = b_i = 0, \label{eqajbi}
\end{equation}
and 
\begin{equation}
   a_i c_m = b_j c_m = c_i a_m + c_j b_m. \label{eqmainkin}
\end{equation}
We prove these relations, which are illustrated in Figure \ref{SI_TreeSwapping} and numerically verified in Figure \ref{SI_MonNumerics}, below.  

After algebraic manipulation, Equation \ref{eqmainkin} implies that
\begin{equation}
\frac{x_m^{ac}}{x_m^{ab}} = \frac{c_j}{a_i} \label{eqfrac1}
\end{equation}
and 
\begin{equation}
\frac{x_m^{cb}}{x_m^{ab}} = \frac{c_i}{a_i}. \label{eqfrac2}
\end{equation}
Because the right hand sides of Equations \ref{eqfrac1} and \ref{eqfrac2} are non-negative, the terms $x^{ac}_m, \ x^{cb}_m$, and $x^{ab}_m$ must share the same sign, proving that coherence conditions are satisfied. 

To establish proportionality of the non-equilibrium responses to perturbations in $F_{ij}$, we apply Equation \ref{eqderivoffset} with $L_m = \theta_0^m$, so that
\begin{eqnarray}
    \frac{\partial \pi_m}{\partial \lambda} &=& \left(\theta_2^m - \theta_0^m \right)\frac{\partial w_2}{\partial \lambda} + \left(\theta_1^m - \theta_0^m \right)\frac{\partial w_1}{\partial \lambda} \nonumber \\
    &=& \frac{x^{ab}_m}{\bar{a}\bar{b}}\frac{\partial w_2}{\partial \lambda} + \frac{x^{cb}_m}{\bar{b}\bar{c}}\frac{\partial w_1}{\partial \lambda} \\ \nonumber 
    &=&x^{ab}_m\left( \frac{1}{\bar{a}\bar{b}}\frac{\partial w_2}{\partial \lambda} + \frac{c_i}{a_i\bar{b}\bar{c}}\frac{\partial w_1}{\partial \lambda}\right). 
\end{eqnarray}
We see that the part depending on node identity $m$ factors from the part depending on $\lambda$, so that this non-equilibrium response is also proportional for all nodes $m$.  

An example of a biochemical process which may modulate an $F_{ij}$ term is a chemical reaction of the form $i + X \leftrightarrow j + Y$ in which the combined amounts of $X$ and $Y$ are conserved.  For instance, one may take ATP concentrations as $X$ and ADP+Pi concentrations as $Y$.  The shift in chemostatted balance of concentrations of $X$ and $Y$ then anti-symmetrically accelerates the forward while decelerating the reverse transition.

\color{black}

\textit{Proof of Equations \ref{eqaibj}-\ref{eqmainkin}.}
Equation \ref{eqaibj} can be established as follows.  The term $a_i$ represents the sum over all directed trees rooted at node $i$ and containing the edge $j \rightarrow i$.  By flipping this edge, each of these trees can be converted into a directed spanning tree rooted at node $j$ and containing the edge $i \rightarrow j$, which is a unique term in the sum $b_j$.  The weights $W_{ij}$ and $W_{ji}$ of the flipped edges are excluded from $a_i$ and $b_j$, respectively, and as a result these terms both represent the weights for the same collection of edge rates corresponding to flows into nodes $i$ and $j$. 
 
Equation \ref{eqajbi} follows from the fact that no directed spanning trees rooted at node $j$ can contain the edge $j\rightarrow i$, and no directed spanning trees rooted at node $i$ can contain the edge $i\rightarrow j$.

\color{\editcolor}
Equation \ref{eqmainkin} can be established as follows.  The first equality is due to Equation \ref{eqaibj}. Inserting the definitions of the quantities in Equation \ref{eqmainkin}, it reads
\begin{widetext}
\begin{align}
\sum_{T_i \in \mathcal{Q}_{j\rightarrow i}} \sum_{T'_m\in \mathcal{Q}_{(-)}} w^-(T_i) w(T'_m) 
 =& \sum_{T_j\in \mathcal{Q}_{i\rightarrow j}} \sum_{T'_m \in \mathcal{Q}_{(-)}} w^-(T_j) w(T'_m) \nonumber \\ 
 =& \sum_{S_m\in \mathcal{Q}_{j\rightarrow i}} \sum_{S'_i\in \mathcal{Q}_{(-)}} w^-(S_m) w(S'_i)  
 + \sum_{U_m\in \mathcal{Q}_{i\rightarrow j}} \sum_{U'_j\in \mathcal{Q}_{(-)}} w^-(U_m) w(U'_j) \label{eqexpsums}
\end{align}
\end{widetext}
where, as before, $\mathcal{Q}_{j\rightarrow i}$ is the set of directed trees containing the directed edge $j \rightarrow i$, $\mathcal{Q}_{i\rightarrow j}$ is the set of directed trees containing the directed edge $i \rightarrow j$, and $\mathcal{Q}_{(-)}$ is the set of directed trees containing neither. 

Let $N_{[ij]}$ denote the number of undirected spanning trees containing the edge $i\leftrightarrow j$ and $N_-$ denote the number of undirected spanning trees without this undirected edge.  Let $N^m_{ij}$ denote the number of directed spanning trees rooted at node $m$ which contain the directed edge $j \rightarrow i$, and similarly for $N^m_{ji}$.  We have $N^m_{ij} + N^m_{ji} = N_{[ij]}$ for any $m$.  The double sums $a_ic_m$ and $b_j c_m$ on the left hand side of Equation \ref{eqexpsums} each have $N_{[ij]} N_-$ terms, because $a_i$ and $b_j$ each have $N_{[ij]}$ (as every instance of $i\leftrightarrow j$ must be directed toward $i$ and $j$, respectively, in these terms), and $c_m$ has $N_-$.  Meanwhile the double sum $c_i a_m$ has $N_-N^m_{ij} $ and $c_jb_m$ has $N_- N^m_{ji} $.  Hence, there are the same number of terms on both sides of the equation.  Referring to the example in Figure \ref{SI_TreeSwapping}, we see that $a_i$, $b_j$, and $c_m$ each contain 4 terms, so that $a_i c_m$ and $b_j c_m$ each contain 16.  On the other hand $a_m$ and $b_m$ each contain 2, and $c_i$ and $c_j$, so that $c_i a_m + c_j b_m$ also contain 16, illustrating this counting argument.

\color{black}

If we can find a one-to-one correspondence between the terms on each side of this equation, the equality would be proved.  Fortunately, just such a correspondence was illustrated in Ref.\ \citenum{owen2020universal} using a procedure called ``tree surgery.''  This procedure algorithmically takes as input a pair of directed spanning trees $T_m$, which is rooted at node $m$ and contains the edge $j \rightarrow i$ or $i \rightarrow j$, and $T'_n$, which is rooted at node $n$ does not contain either edge.  The procedure converts this pair into a new pair $\tilde{T}_n$ and $\tilde{T}'_m$, where $\tilde{T}_n$ contains the edge $j \rightarrow i$ or $i \rightarrow j$, and $\tilde{T}'_m$ does not.  Two examples of this are illustrated in the example in Figure \ref{SI_TreeSwapping}.  It was further shown in Ref.\ \citenum{owen2020universal} that the weight products $w(T_m) w(T'_n)$ and $w(\tilde{T}_n) w(\tilde{T}'_m)$ will be equal if, during the conversion procedure, the edge $j \rightarrow i$ or $i \rightarrow j$ is not flipped. 

Using this procedure, it is possible to convert each of the $N_{[ij]} N_-$ terms on the left hand side of Equation \ref{eqexpsums} into either a term of the form $w^-(S_m) w(S'_i)$, or one of the form $w^-(U_m) w(U'_j)$.  In our case, one can ignore the effect of flipping the distinguished edge $i \leftrightarrow j$ because doing so would simply convert a tree $T_i$ rooted at $i$ into a tree $T_j$ rooted at $j$, or vice versa.  The weights of interest $w^{-}(T_i) = w^{-}(T_j)$ for two such trees must be the same because the term $W_{ij}$ in $w(T_i)$ and $W_{ji}$ in $w(T_j)$, which would change during the edge flip, have been factored out in $w^-(T_i)$ and $w^-(T_j)$.  As a result, flipping the edge $j \rightarrow i$ or $i \rightarrow j$ does not affect the equality of edge weights between the left and right hand sides of Equation \ref{eqexpsums}.  We can therefore establish the desired one-to-one correspondence between the terms on both sides of Equation \ref{eqexpsums}, implying equality. 

\color{\editcolor}
\section{Equivalence of coherence conditions with previous analyses of transcription regulation models}\label{App:Incoherence}

Here we consider the model in Ref.~\citenum{martinez2025emergence} for regulation of DNA transcription and show that our general sufficient conditions for monotonicity, Equations~\ref{eqmonorderinc} and~\ref{eqmonorderdec}, reduce to the conditions derived in that study for monotonic response functions.  The model is illustrated in Figure~\ref{BiologicalExamples}A, ignoring here the feedback from the GP to the TF binding transitions. The observable of interest is a linear combination of the set of states with RNAP bound, $C = \pi_3(\lambda) + \epsilon \pi_4(\lambda),$ viewed as a function of the TF concentration $\lambda$ which linearly scales the first-order rates $W_{43}$ and $W_{21}$. The rates are also parameterized through $\epsilon_a = W_{42}/W_{31}$, $\epsilon_b = W_{13}/W_{24}$, and $\omega = \epsilon_a / \epsilon_b$.

We note that in equilibrium models of DNA transcription regulation, the cycle condition holds for any value of $\lambda$, and the steady-state construction yields an expression for $C$ of the form of Equation~\ref{eqCdef} with $N_\lambda = 1$ (Equation~4 in Ref.~\citenum{martinez2025emergence}), so that $C(\lambda)$ is automatically monotonic. More generally, although Equation~\ref{eqpimlambda} arises naturally from the matrix-tree theorem for non-equilibrium steady states, expressions of the same form also appear in graphical constructions for equilibrium systems \cite{martinez2025emergence, estrada2016information}. In several specific equilibrium models we find that the resulting distributions satisfy the coherence conditions introduced here, providing a natural mathematical explanation for the monotonic response behavior observed in those models. For example, in the ``all-or-nothing'' model of equilibrium transcriptional regulation studied in Ref.~\citenum{estrada2016information}, the protein expression as a function of TF concentration $[\text{TF}]$ is proportional to
\begin{equation}
   C([\text{TF}])=\frac{c_{n_T}[\text{TF}]^{n_T}}{\sum_{n=0}^{n_T} c_{n}[\text{TF}]^{n}}
\end{equation}
with $n_T$ the number of TF binding sites and the $c_n$ all positive, which clearly satisfies Equation~\ref{eqmonorderinc}.  As another example, in the ``average-binding'' model, the expression is given by  
\begin{equation}
   C([\text{TF}])=\frac{\sum_{n'=1}^{n_T} (n'/n_T)c_{n'}[\text{TF}]^{n'}}{\sum_{n=0}^{n_T} c_{n}[\text{TF}]^{n}}
\end{equation}
which also satisfies Equation~\ref{eqmonorderinc}.  More broadly, this shows that the model-specific mechanisms identified in these earlier studies are particular cases of the general structural criteria in Equations~\ref{eqmonorderinc} and \ref{eqmonorderdec}, in which monotonicity is governed by network-level coherence conditions that apply to arbitrary Markov processes with rate modulation.

Outside of the equilibrium regulation regime, the expression for $C(\lambda)$ is of the form of Equation~\ref{eqCdef} with $N_\lambda = 2$, so that monotonicity is no longer guaranteed as it is when $N_\lambda = 1$. In Ref.~\citenum{martinez2025emergence}, the conditions for monotonicity are established through algebraic inspection of the derivative $\partial C / \partial \lambda$ (Equations~S8-S10 in their SI Appendix). A systematic alternative to this approach by inspection is to evaluate the coherence conditions, Equations~\ref{eqmonorderinc} and~\ref{eqmonorderdec}. This can be done by determining whether the sign of $\zeta^C_{\mu+1}\bar{\zeta}_\mu - \zeta^C_{\mu}\bar{\zeta}_{\mu+1}$ is the same for all $\mu$. For the model considered here, doing so yields two expressions whose signs are determined by sums of quantities proportional to $\epsilon-1$, $\epsilon_a - 1$, $\epsilon_b - 1$, $\epsilon_b \epsilon - 1$, and $\epsilon_a \epsilon_b \epsilon - 1$. A sufficient condition for $C(\lambda)$ to be monotonic is therefore that $\epsilon_a$, $\epsilon_b$, and $\epsilon$ are either all less than one or all greater than one, which matches the result obtained by inspection in Ref.~\citenum{martinez2025emergence}.  This result also generalizes the finding in Equation 5 of Ref.~\citenum{mahdavi2024flexibility} for $\epsilon = 1$.  We note that for our parameters, provided in Appendix \ref{App:NumMeth}, we have $\epsilon_a >1$ and $\epsilon_b>1$.

The coherence conditions discussed in the present paper thus provide a systematic route to determine the algebraic regimes in which monotonicity can occur for models with $N_\lambda > 1$.  Although the resulting expressions may not be easily interpretable in complex models, they can be numerically evaluated  (as in the models of \textit{E. coli} chemotaxis Figure~\ref{Ladders}).

\color{black}

\begin{figure*}[ht!]
\begin{center}
\includegraphics[width=\textwidth]{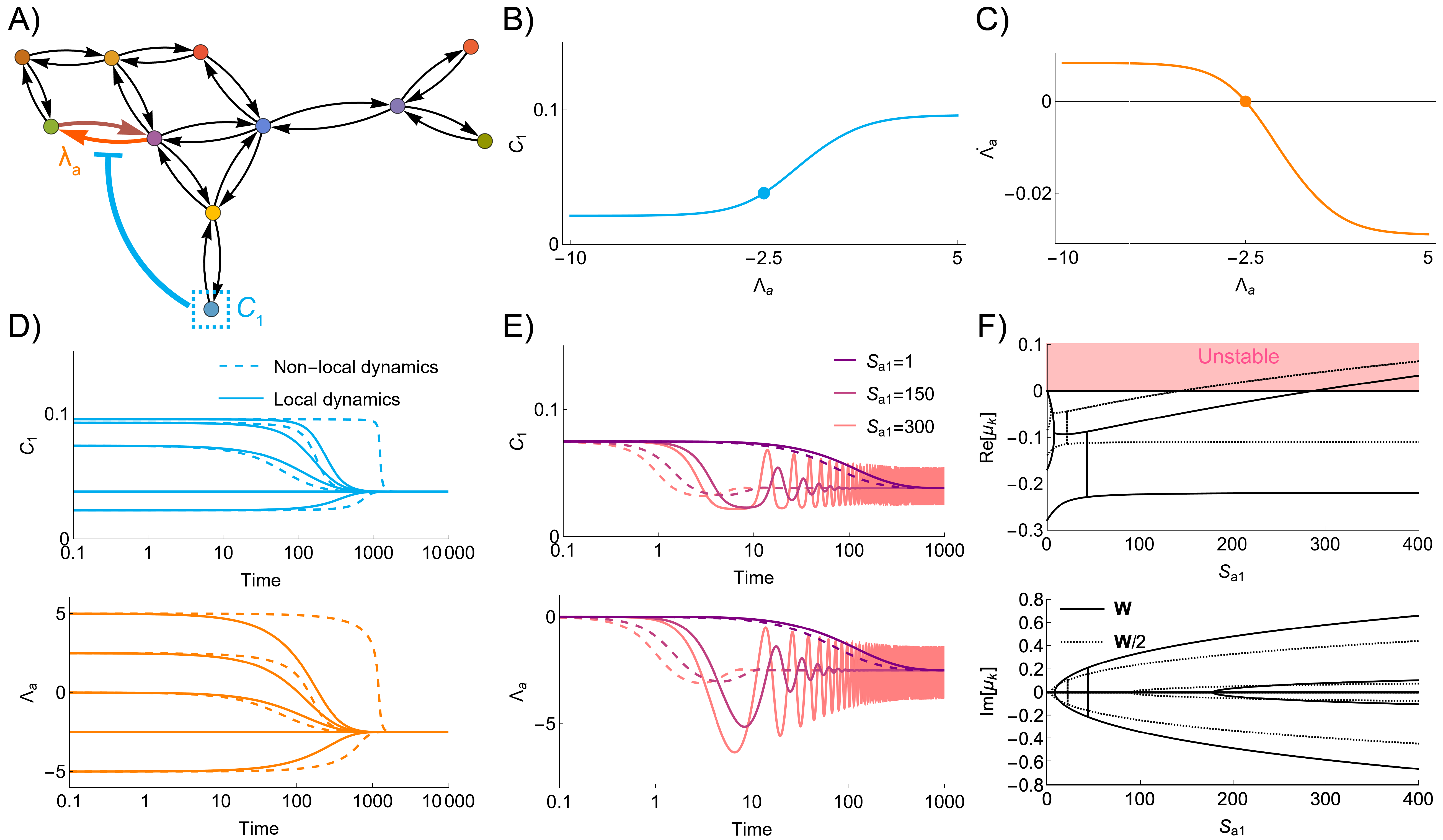}
\caption{\textbf{One-dimensional learning dynamics.} 
A) The graph system used in this example with the observable $C_1$ outlined, with randomly sampled edge rates (see Appendix \ref{App:NumMeth} and Figure~\ref{2DLearning}). The edge perturbation $\lambda_a=\exp F_{ij}/2$ is applied at the orange transitions. 
B)  Plot of the non-equilibrium adiabatic response $C_1(\bp(\Lambda_a)) \equiv C_1(\Lambda_a)$ where $\Lambda_a = \ln \lambda_a$, and with the target value at $C_1^\text{env} = C_1(-2.5)$ shown as a dot. 
C) Phase plot of the adiabatic dynamics $\dot{\Lambda}_a(\Lambda_a) = S_{a1}(C_1(\Lambda_a) - C_1^\text{env})$ with $S_{a1} = 1$, showing global stability of these feedback dynamics.  
D) \textit{Top}: Trajectories of $C_1$ under the adiabatic local dynamics $\dot{\Lambda}_a(\Lambda_a) = S_{a1}(C_1(\Lambda_a) - C_1^\text{env})$ from several initial conditions of $\Lambda_a(t=0)$.  Also plotted are trajectories from the adiabatic non-local dynamics $\dot{\Lambda}_a(\Lambda_a) = -(\partial C_1(\Lambda_a) / \partial \Lambda_a)(C_1(\Lambda_a) - C_1^\text{env})$ which are always guaranteed to be globally stable.  \textit{Bottom}: Trajectories of $\lambda_a$ corresponding to the top plots.  
E) For the same graph system, plots of trajectories of the non-adiabatic dynamics $\dot{\Lambda}_a(\Lambda_a) = -S_{a1}(C_1(\mathbf{p}) - C_1^\text{env})$, with $\mathbf{p}(t)$ used instead of $\bp(\Lambda_a(t))$, for increasing values of $S_{a1}$.  Dashed lines show the non-local non-adiabatic dynamics $\dot{\Lambda}_a(\Lambda_a) = -S_{a1} (\partial C_1 / \partial \Lambda_a)(C_1 - C_1^\text{env})$.  For the largest value of $S_{a1}$, there are oscillatory instabilities for the local dynamics but not for the non-local dynamics.   F) Real part of the eigenspectrum of $\mathbf{G}$ the combined system $(\mathbf{p},\dot{\Lambda}_a) = \mathbf{g}(\mathbf{p},\Lambda_a)$ (see Equations~\ref{eqgna} and \ref{eqGna}) as $S_{a1}$ is increased.  An instability occurs when branches of the eigenspectrum take on positive real parts, indicated by the pink region.  The dotted lines show the eigenspectrum when the rate matrix elements have been halved, thus making the feedback dynamics comparatively faster and the onset of instability lower.  \textit{Bottom}:  Same as the top, but for the imaginary parts of the eigenspectra.}
\label{SI_TimescaleInstability}
\end{center}
\end{figure*}

\section{General formulation of feedback dynamics}\label{App:FormFeed}

Here we study the general form of the local imperfect feedback dynamics, and we compare it with non-local feedback dynamics which are ensured to be globally stable.  We first consider the adiabatic limit, in which the general dynamics for $\bL\in \mathbb{R}^{D}$ taking feedback from the network observables can be written  
\begin{equation}
   \dot{\bL} = \mathbf{M}(\bL)(\mathbf{C}(\bp(\bL)) - \mathbf{C}^\text{env}), \label{eq:adot}
\end{equation}
where $\mathbf{C} \in \mathbb{R}^{N_\text{C}}$ is a vector of network observables, and $\mathbf{M} \in \mathbb{R}^{D \times N_\text{C}}$ is a matrix that will be specified below.  For brevity we write $\mathbf{C}(\bL) \equiv \mathbf{C}(\bp(\bL))$ in the adiabatic limit  and define the Jacobian matrix  
$\mathbf{J}(\bL) \equiv \nabla_{\bL} \mathbf{C}(\bL) \in \mathbb{R}^{D \times N_\text{C}}.$  Assuming that $\mathbf{C}^\text{env}$ is reachable as $\mathbf{C}(\bL^\text{env})$ for some $\bL^\text{env}$, we linearize these dynamics around the fixed point as 
\begin{equation}
\dot{\bL} \approx \mathbf{M}(\bL^\text{env})\mathbf{J}^\top(\bL^\text{env})\left(\bL - \bL^\text{env} \right).
\end{equation}  
Local stability thus depends on the eigenvalues of $\mathbf{M}(\bL^\text{env})\mathbf{J}^\top(\bL^\text{env})$.  

One could ensure global stability of Equation~\ref{eq:adot} by choosing $\mathbf{M}(\bL) = -\mathbf{J}(\bL)$. This choice is non-local because it requires knowledge of the Jacobian elements $J_{a1} = \frac{\partial C_1}{\partial \Lambda_a},$
which involve rate matrix parameters across the entire network and which may additionally vary with $\bL$. Under this choice, we can define the Lyapunov function  $\mathcal{L} = \frac{1}{2} \|\mathbf{C} - \mathbf{C}^\text{env}\|^2,$ whose gradient with respect to $\bL$ is  
$\nabla_{\bL} \mathcal{L} = \mathbf{J}(\bL)(\mathbf{C}(\bL) - \mathbf{C}^\text{env})$.   Equation~\ref{eq:adot} with $\mathbf{M}(\bL) = -\mathbf{J}(\bL)$ thus flows down the gradient of a Lyapunov function and is hence globally stable.  

We can approximate these non-local dynamics by choosing $\mathbf{M}(\bL) = \mathbf{S} \sim -\mathbf{J}^\top(\bL^\text{env})$, where $\mathbf{S}$ is a sparse constant matrix. This choice is local, because it only couples a fraction of observables to the feedback variables, and the constant matrix $\mathbf{S}$ does not involve network-level and $\bL$-dependent information.  The local stability of the approximate dynamics at the steady state is determined by the eigenvalues of the matrix $ \mathbf{S} \mathbf{J}^\top(\bL^\text{env}) \in \mathbb{R}^{D\times D}$. This matrix has rank at most $D$, and we may ask the conditions for all $D$ eigenvalues to be negative. This requires two conditions.  First, $ \mathbf{S} $ must have full rank, which may fail if two observables redundantly provide feedback to two separate regulated edges (see Figure \ref{2DLearning}).  Second, at least $D$ columns of $ \mathbf{J}^\top(\bL^\text{env}) $ must not lie in the kernel of $ \mathbf{S} $. This condition relates to whether the sparse projection of the network observables, $ \mathbf{C}(\bL^\text{env}) $, can be ``inverted'' to uniquely determine $ \bL^\text{env} $. While the forward problem, solving for $ \mathbf{C}(\bp(\bL))$ given $ \bL $, has a unique solution for any ergodic graph and any $\bL$, the inverse problem of determining $ \bL $ from observed steady-state node values $\mathbf{C}$ can only be uniquely solved for certain choices of network observables. We discuss this further in Appendix \ref{App:UniqueFixed}.  If network observables are chosen such that they do not allow for uniquely solving $ \bL(\mathbf{C}) $, and these are used as the nonzero elements in $ \mathbf{S} $, then $ \mathbf{S} \mathbf{J}^\top(\bL^\text{env}) $ will have rank less than $ D$, compromising stability.  

In Figure \ref{SI_TimescaleInstability} we compare these local and non-local dynamics for a randomly sampled graph.  We first show in the adiabatic limit for $D=1$ that both dynamics are globally stable.  This is true for the non-local dynamics because they minimize a Lyapunov function, and it is true for the local dynamics by virtue of the thermodynamic constraints we have derived saying that non-equilibrium response signs cannot flip when $D=1$.  
 
Outside of the adiabatic limit, the local feedback dynamics are given by
\begin{equation}
\begin{pmatrix}
\dot{\mathbf{p}} \\ \dot{\bL} 
\end{pmatrix} = \begin{pmatrix}
\mathbf{W}(\bL)\mathbf{p} \\ \mathbf{S}\left(\mathbf{C}(\mathbf{p}) - \mathbf{C}^\text{env} \right)  \end{pmatrix} \equiv \mathbf{g}(\mathbf{p},\bL). \label{eqgna}
\end{equation}
The stability of this system at the fixed point $(\bp(\bL^\text{env}), \bL^\text{env})$ depends on the eigenvalues of the matrix
\begin{equation}
    \mathbf{G} = \nabla_{(\mathbf{p},\bL)}\mathbf{g}\left(\bp(\bL^\text{env}), \bL^\text{env}\right). \label{eqGna}
\end{equation}
In Figure \ref{SI_TimescaleInstability}E-F we show in a one-dimensional example that when the feedback $S_{a1}$ is very large in magnitude, oscillatory steady states may arise instead of fixed points.  This phenomenon is analogous to oscillatory instabilities observed in other control systems with excessive feedback gain \cite{bechhoefer2021control, del2015biomolecular}, and the onset of this instability coincides with the appearance of real positive parts in the eigenspectrum of $\mathbf{G}$.  

\color{\editcolor}
Quantitative predictions for this critical gain are difficult, due to the complex interplay of Markovian timescales which can sometimes produce counter-intuitive relaxation dynamics \cite{lu2017nonequilibrium}, and understanding how critical feedback strengths in the non-adiabatic regime relate to the topological structure of the Markov graph and the choice of regulated edges remains an open and important problem. We expect that an ``optimal'' gain corresponds to a critically damped response, allowing for rapid response times without oscillations. In Appendix \ref{App:Non-adia} we derive the first-order correction to the adiabatic approximation, but this perturbative approach does not make the onset of instability analytically accessible, since the first-order approximation typically breaks down before the loss of dynamical stability; a more complete treatment would likely require a global analysis of the combined network-feedback dynamics, accounting for the full spectrum of relaxation timescales and their dependence on network topology. One possible direction for future progress on the non-adiabatic problem is through connections with works on 
counterdiabatic driving \cite{ilker2022shortcuts, patra2017shortcuts, iram2021controlling}, which provide systematic methods for controlling transitions between states at finite speeds and could offer tools for characterizing how finite-timescale corrections to the adiabatic dynamics interact with feedback gain, potentially enabling tighter bounds on the critical gain as a function of network topology. In most of this paper we work in the adiabatic limit for simplicity; in the context of DNA regulation, this implies the realistic approximation that transcription of GP and its regulation of TF or DNA kinetics be slower than the local binding kinetics of TF and RNAP on DNA. 
\color{black}

\color{\editcolor}
\section{Non-adiabatic correction to the feedback dynamics}\label{App:Non-adia}

Here we derive a first-order correction to the adiabatic approximation of the feedback dynamics.  We note that such non-adiabatic dynamics of Markov processes have previously been studied in the open-loop control setting, in which for example certain geometric pumping theorems have been obtained \cite{rahav2008directed, takahashi2020nonadiabatic}.  Here, instead, the control variable $\bL$ is dynamically coupled to the system state $\mathbf{p}$ through the local imperfect feedback connections, in a closed-loop setting \cite{bechhoefer2021control}.  

To explicitly identify the relevant timescales, we write the feedback dynamics as (c.f. Equation~\ref{eqgna})
\begin{eqnarray}
    \tau_p \dot{\mathbf{p}}(t) &=& \mathbf{W}(\bL(t)) \mathbf{p}(t), \label{eqpdotfullSI} \\
    \tau_\lambda \dot{\bL}(t) &=& \mathbf{S} \mathbf{p}(t) - \bL_0 \label{eqbldynamics}
\end{eqnarray}
where we have introduced $\tau_p$ and $\tau_\lambda$ by scaling the kinetic terms on the right-hand sides.  We have also absorbed into $\mathbf{S}$ any weights $q_m$ for the definition of the observable $\mathbf{C}(\mathbf{p})$ (c.f. Equation~\ref{eqCdefOrig}), and we absorbed into $\bL_0$ all constant terms in the dynamics for $\bL$.  

In the adiabatic limit $\tau_p \rightarrow 0$ with finite $\tau_\lambda$, in which the Markov process $\mathbf{p}$ is always at steady state with respect to $\bL(t)$, we have
\begin{eqnarray}
    \mathbf{p}(t) &=& \bp(\bL(t)), \\
    \tau_\lambda \dot{\bL}(t) &=& \mathbf{S} \bp(\bL(t)) - \bL_0, \label{eqbldynamicsad}
\end{eqnarray}
where $\bp(\bL)$ satisfies $\mathbf{W}(\bL)\bp(\bL)=\mathbf{0}$ and $\mathbf{1}^\top\bp(\bL)=1$.

To go beyond the adiabatic limit we next find the first-order correction for the $\bL$ dynamics in the small parameter $\tau_p / \tau_\lambda$.  We write the deviation from the adiabatic trajectory by introducing the vector $\brho(t)$ (satisfying $\mathbf{1}^\top\brho(t)=0$), as
\begin{equation}
    \mathbf{p}(t) = \bp(\bL(t)) + \tau_p \brho(t),
\end{equation}
so that Equation~\ref{eqpdotfullSI} reads
\begin{equation}
    \tau_p \left(\nabla_\bL \bp \right)\dot{\bL}(t) 
    + \tau_p^2 \dot{\brho}(t)
    = \mathbf{W}(\bL(t)) \left( \bp(\bL(t)) + \tau_p \brho(t)\right).
\end{equation}
Rearranging, and using $\mathbf{W}(\bL)\bp(\bL)=\mathbf{0}$, we obtain to lowest order in $\tau_p$
\begin{equation}
    \left(\nabla_\bL \bp \right)\dot{\bL}(t) 
    = \mathbf{W}(\bL(t))\brho(t).
\end{equation}

To solve for $\brho(t)$, which is a zero-sum vector living on the tangent space to the probability simplex, we can restrict $\mathbf{W}$ to this subspace and invert it.  This yields the Drazin pseudo-inverse $\mathbf{W}^+(\bL)$, with which we write
\begin{equation}
    \brho(t) 
    = \mathbf{W}^+(\bL(t)) 
      \left(\nabla_\bL \bp \right)\dot{\bL}(t).
\end{equation}
Substituting this expression into the $\bL$ dynamics (Equation \ref{eqbldynamics}) gives
\begin{equation}
    \tau_\lambda \dot{\bL}
    = \mathbf{F}(\bL)
    + \tau_p \mathbf{G}(\bL) \dot{\bL}, \label{eqbldynamics2}
\end{equation}
where
\begin{eqnarray}
    \mathbf{F}(\bL) &\equiv& \mathbf{S}\bp(\bL) - \bL_0, \\
    \mathbf{G}(\bL) &\equiv& \mathbf{S}\mathbf{W}^+(\bL) \left(\nabla_\bL \bp(\bL)\right).
\end{eqnarray}

Rearranging Equation \ref{eqbldynamics2} gives the dynamics
\begin{equation}
    \left(\mathbf{I} - \beta \mathbf{G}(\bL)\right)\dot{\bL}
    = \frac{1}{\tau_\lambda}\mathbf{F}(\bL),
\end{equation}
where
\begin{equation}
      \beta \equiv \tau_p/\tau_\lambda,
\end{equation}
or
\begin{equation}
    \dot{\bL}
    = \frac{1}{\tau_\lambda}
    \left(\mathbf{I} - \beta \mathbf{G}(\bL)\right)^{-1}
    \mathbf{F}(\bL).
    \label{eqfirstoder}
\end{equation}
This reduces to the purely adiabatic dynamics, Equation \ref{eqbldynamicsad}, when $\beta  = 0$.  In Figure \ref{SI_Nonadia} we show these first-order non-adiabatic dynamics compared to the full dynamics and the adiabatic dynamics when $\beta = 0.5$.  

This analysis focuses primarily on the interplay of the timescales $\tau_p$ and $\tau_\lambda$. The instantaneous eigenspectrum of $\mathbf{W}(\boldsymbol{\lambda}(t))$, particularly its slow modes, introduces additional relative timescales that govern both the accuracy of the adiabatic approximation and the onset of instability from feedback gain. This eigenspectrum reflects the full Markov graph structure, and the interplay among eigenmodes with non-zero initial amplitudes can produce non-trivial relaxation dynamics \cite{lu2017nonequilibrium}. Fully characterizing how the graph structure and its eigenmodes determine the validity of the adiabatic approximation is thus an open problem which we leave for future work.

\begin{figure}
\begin{center}
\includegraphics[width=\columnwidth]{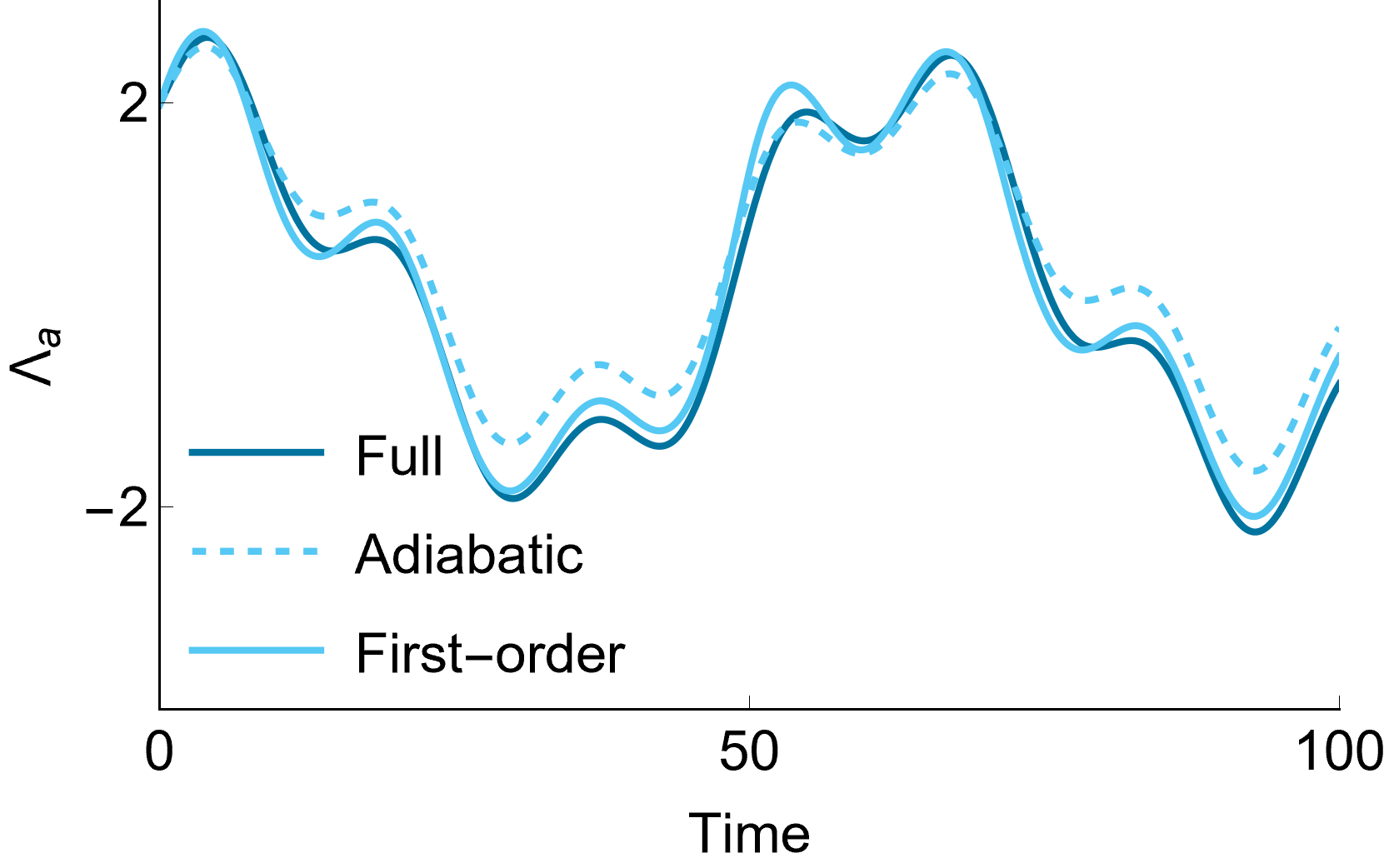}
\caption{
\color{\editcolor}
\textbf{Non-adiabatic correction to the feedback dynamics.}  For the feedback dynamics studied in Figure \ref{BiologicalExamples}C and with $\beta = 0.5$, we show the gain in accuracy allowed by the first-order non-adiabatic correction (Equation \ref{eqfirstoder}) compared to the fully adiabatic approximation (Equation \ref{eqbldynamicsad}).  Here $\Lambda_a = \Delta E_1$.        
\color{black}
}
\label{SI_Nonadia}
\end{center}
\end{figure}

\color{black}

\begin{figure*}
\begin{center}
\includegraphics[width=0.85\textwidth]{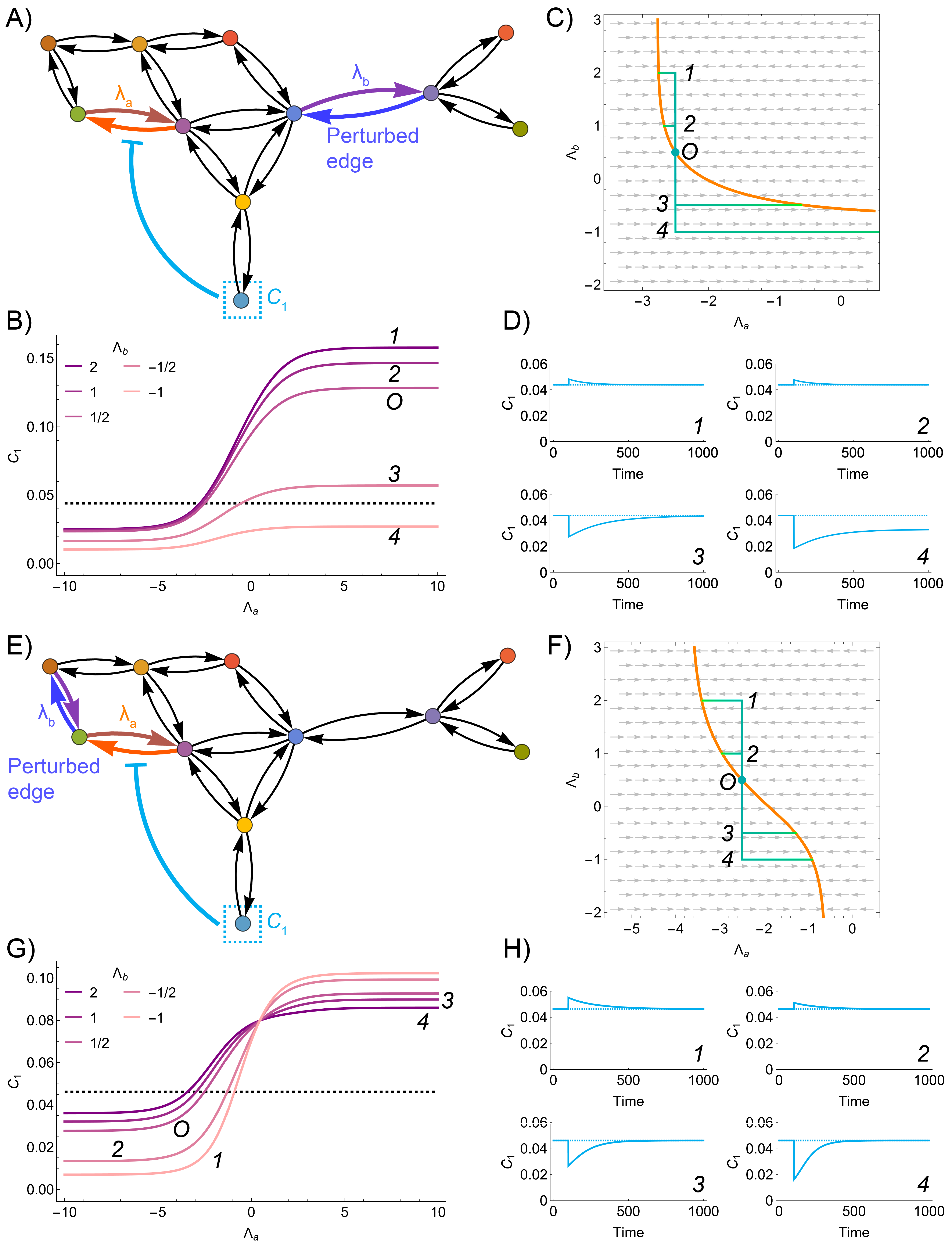}
\caption{\textbf{Adaptation to changes in network parameters dependent on reachability.}  A) The graph system used in this example; c.f. Figures~\ref{2DLearning} and \ref{SI_TimescaleInstability}.  The variable $\Lambda_b = F_{kl}/2 $ along the blue edge is perturbed as a step change in the remaining panels.  B) Plots of the non-equilibrium response $C_1(\bp(\Lambda_a;\Lambda_b))$ for various values of the perturbation in $\Lambda_b$.  The horizontal dashed line indicates the value $C_1(-2.5;0.5)$, which all curves except $C_1(\Lambda_a;-1)$ intersect and can hence reach through variation in $\Lambda_a$.  C)  Phase plane and trajectories for the one-dimensional adiabatic dynamics $\dot{\Lambda}_a = S_{a1}(C_1(\Lambda_a;\Lambda_b) - C_1^\text{env})$ with $S_{a1} = 1$ and fixed $C_1^\text{env}$, and with various step changes in the perturbed $\Lambda_b$.  For cases 1, 2, and 3, the dynamics flow back to the nullcline, indicating perfect adaptation under the new parameter $\Lambda_b$, but for case 4 the flow does not intersect the nullcline.  D)  Same as panel C, but showing the trajectories of $C_1$. E-H) Same as panels A-D, for a graph in which, due to the structure of the nullclines, perfect adaptation to any change in $\Lambda_b$ is possible through feedback on $\Lambda_a$.  }
\label{SI_Adaptation}
\end{center}
\end{figure*}

\color{\editcolor}
\section{Adaptation and reachability}\label{App:Adaptation}

In Figure \ref{SI_Adaptation} we show an example of a network which perfectly adapts only for a certain range of perturbation sizes, and an example of a network which perfectly adapts for any perturbation size. As discussed in the main text, the ability of a network to adapt to a perturbation depends on whether the set point remains reachable through changes in the regulated edges after the perturbation. Characterizing reachability analytically in terms of interpretable features of the network topology is difficult, as the range of the observable depends nonlinearly on the transition rates in a way that reflects the entire graph structure. We thus leave a full analytical investigation to future work, but we note related work that may prove useful in this direction \cite{chittari2023geometric}.  We emphasize, however, that, provided reachability of the target point, the ability of steady states to adapt with arbitrary network topologies using simple linear feedback rules on any regulated edges can be contrasted with previous works which consider specially designed motifs or network substructures, or require various assumptions about the system's response \cite{lan2012energy, tu2018adaptation, barkai1997robustness, alon2019introduction, araujo2018topological, araujo2023universal, ma2009defining, aoki2019universal, PRXLife.3.013017, briat2016antithetic}.
\color{black}

\section{Unique fixed points}\label{App:UniqueFixed}
Here we consider solutions to the steady-state equation 
\begin{equation}
    \mathbf{W}(\bl)\bp = \mathbf{0} \label{eqmatrixeq}
\end{equation}
where $\bl \in \mathbb{R}^{D}$, and we assume that we observe a set of steady-state probabilities $\{\pi_k\}_{k \in \mathcal{K}}$ where $\mathcal{K}$ is a set of indices of size $D$.  For an ergodic network there is a unique steady state $\bp$ as a function of $\mathbf{W}(\bl)$, so the mapping from $\bl$ to $\tilde{\bp} \in \mathbb{R}^D$ (the vector with elements $\{\pi_k\}_{k \in \mathcal{K}}$) is a function.  We now ask if we can uniquely invert this function.  

Equation \ref{eqmatrixeq} represents $N_\text{n}$ constraints on the $D + N_\text{n}$ variables in $\bl$ and $\bp$, but these constraints are not all independent as the columns of $\mathbf{W}$ sum to zero.  We can supplement these equations with the normalization condition $\sum_m \pi_m = 1$ to provide $N_\text{n}$ independent constraints.  The typical forward problem is that, knowing $\bl$, we use these $N_\text{n}$ constraints to determine the steady-state vector $\bp$.  We now consider a partial inverse problem: knowing $\tilde{\bp} \in \mathbb{R}^{D}$, we want to determine the $N_\text{n} - D$ remaining elements of $\bp$ and the $D$ elements of $\bl$.  It seems that we should be able to do this because we have enough independent constraints, but for some choices of rate perturbations the elements of $\bl$ enter non-linearly into $\mathbf{W}(\bl)$ so we need to check if solutions are unique.  

We first consider the case $D = 1$, meaning there is one regulated edge parameter $\Lambda_a = \ln \lambda_a$.  If $\Lambda_a$ corresponds to rate perturbations of the types $X_{ij}, \ E_i,$ or $B_{ij}$, then it is straightforward to show that the Equations \ref{eqmatrixeq} are linear in $e^\Lambda_a$ so any inverse solutions with enough independent constraints are unique.  If $\Lambda_a$ corresponds to $F_{ij}$-type perturbations, then the resulting equations are quadratic and take more consideration.  In this case the driving force $F_{ij}$ is in the direction $j\rightarrow i$, so that $W_{ij} = W_{ij}^0e^{\Lambda_a / 2}$ and $W_{ji} = W_{ji}^0 e^{-\Lambda_a/2}$.  We first reserve the $i$ and $j$ rows of Equation \ref{eqmatrixeq} and use the remaining $N_\text{n} -2$ independent constraints to solve for $N_\text{n}-2$ of the $N_\text{n} -1$ unknown elements of $\bp$, leaving one unknown element remaining.  It can be shown that the $i^\text{th}$ row of Equation \ref{eqmatrixeq} has the form
\begin{equation}
    W_{ij}^0 \pi_j e^{\Lambda_a / 2} - W_{ji}^0\pi_i e^{-\Lambda_a / 2} + Q_i = 0 \label{eqc1}
\end{equation}
where $Q_i$ collects quantities which do not depend on $\Lambda_a$.  The $j^\text{th}$ row has a similar form
\begin{equation}
    -W_{ij}^0 \pi_j e^{\Lambda_a / 2} + W_{ji}^0\pi_i e^{-\Lambda_a / 2} + Q_j = 0. \label{eqc2}
\end{equation}
We can now add Equation \ref{eqc1} to Equation \ref{eqc2} to yield $Q_i = - Q_j$, which provides the final linear constraint needed to fix the last unknown element of $\bp$.  Setting $Q_i = -Q_j$, we eliminate one of the above two equations and treat the remaining one as a quadratic function in the unknown quantity $x_a = e^{\Lambda_a/2} > 0$:
\begin{equation}
    W_{ij}^0 \pi_j x_a^2 + Q_i x_a - W_{ji}^0 \pi_i = 0.
\end{equation}
Regardless of the sign of $Q_i$, this quadratic equation has at most $1$ change in signs between its ordered coefficients, because the coefficient of $x_a^2$ is positive while the coefficient of $x_a^0$ is negative.  By Descartes' rule of signs, this equation thus has at most one positive solution for $x_a$.  We have therefore found unique solutions for all of the unknown variables.    

In the case $D > 1$ the argument proceeds as before, and the only new difficulty arises if two regulated edges are incident on the same node.  To illustrate this we consider the case $D = 2$ with regulated edges $j\rightarrow i$ and $k\rightarrow i$.  Reserving rows $i,j$, and $k$ of Equation \ref{eqmatrixeq} we can use the $N_\text{n}-3$ independent constraints to solve for all but one of the unknown elements in $\bp$.  The $i,j,k$ rows then read, respectively
\begin{eqnarray}
    W_{ij}^0 \pi_j e^{\Lambda_a / 2} - W_{ji}^0\pi_i e^{-\Lambda_a / 2} + Q_i(\Lambda_b) &=& 0, \label{eqc1v}\\
     -W_{ij}^0 \pi_j e^{\Lambda_a / 2} + W_{ji}^0\pi_i e^{-\Lambda_a / 2} - Q_j(\Lambda_b) &=& 0, \label{eqc2v} \\
     -W_{ik}^0 \pi_k e^{\Lambda_b / 2} + W_{ki}^0\pi_i e^{-\Lambda_b / 2} + Q_k(\Lambda_a) &=& 0. \label{eqc3v}
\end{eqnarray}
We begin by combining Equations~\ref{eqc1v} and \ref{eqc2v}, yielding $Q_i(\Lambda_b) = -Q_j(\Lambda_b)$. This relation allows us to solve for the final unknown component of $\boldsymbol{\pi}$. Substituting back into Equation~\ref{eqc1v}, we then solve for the unique positive value of $\Lambda_a$ as a function of $\Lambda_b$. Finally, we use Equation~\ref{eqc3v} to determine the unique positive solution for $\Lambda_b$. 

In the arguments above we assumed that the $N_{\text{n.r.}}$ non-reserved rows (which do not involve the regulated edges) can be used to uniquely eliminate $N_{\text{n.r.}}$ degrees of freedom from the $N_\text{n} - D$ unknown variables.  However, this will not be the case if those rows are not all independent, having low rank.  Uniquely inverting the mapping $\bl \rightarrow \tilde{\bp}$ thus depends on the specific graph and the choice of $\mathcal{K}$ and the regulated edges.  

\section{Numerical methods}\label{App:NumMeth}

The steady-states of the Markov jump process were obtained using the matrix-tree theorem.  We used Mathematica's IGraph library to enumerate the directed spanning spanning trees of the reaction graph \cite{horvat2023igraph}.  The dynamical trajectories were generated using Mathematica's NDSolve function. 

To sample random graphs we draw from the uniform distribution over graphs with $N_\text{n}$ nodes and $N_\text{e}$ edges.  We initialize the edge rates $W_{ij}$ using Equation \ref{eqmarkovrate} by drawing the $N_\text{n} + 2 N_\text{e}$ parameters $E_i$, $B_{ij}$, $F_{ij}$ from uniform distributions over the ranges $[-E_R,E_R]$, $[-B_R,B_R]$, and $[-F_R,F_R]$ respectively.  Unless noted otherwise we take $E_R = B_R = F_R = 1$ throughout the paper.  

\color{\editcolor}
We study throughout the four-state transcriptional model analyzed in Ref.~\citenum{martinez2025emergence}; see Figure \ref{BiologicalExamples}A.  For this system we use the matrix
\begin{widetext}
\begin{equation}
    \mathbf{W} = \begin{pmatrix}
-W_{21}e^{\Delta E_1}[\text{TF}] - W_{31} [\text{RNAP}]e^{\Delta E_1} & W_{12} & W_{13} & 0 \\
W_{21}e^{\Delta E_1}[\text{TF}] & -W_{42}[\text{RNAP}] - W_{12} & 0 & W_{24} \\
W_{31}e^{\Delta E_1}[\text{RNAP}] & 0 & -W_{13} - W_{43}[\text{TF}] & W_{34} \\
0 & W_{42}[\text{RNAP}] & W_{43}[\text{TF}] & - W_{24} - W_{34} 
\end{pmatrix}.
\end{equation}
\end{widetext}
All regulation variables $\Delta E_1, \ [\text{RNAP}]$, and $[\text{TF}]$ are explicitly shown here, although not all are simultaneously varied.  Unless otherwise specified, we follow the paramaterization in Ref.~\citenum{martinez2025emergence} and set $W_{12} = 1, \ W_{13} = 0.1, \ W_{21} = 0.01, \ W_{24} = 0.01, \ W_{31} =  0.01, \ W_{34} = 1, \ W_{42} = 0.1$, and $W_{43} = 0.01$, and we set by default $\Delta E_1 = 0$, $[\text{RNAP}] = 1$, and $[\text{TF}] = 1$.  

\color{black}

\section*{Acknowledgments}
We wish to thank Jordan Horowitz, Menachem Stern, Agnish Kumar Behera, Matthew Du, Billie Meadowcroft, and Nicolas Romeo for helpful discussions.  This work was supported by the National Institute of General Medical Sciences of the NIH under Award No. R35GM147400 by funding to SV and CF.  We also acknowledge funding from the Physics Frontier Center for Living Systems funded by the National Science Foundation (PHY-2317138). CF acknowledges support from the University of Chicago through a Chicago Center for Theoretical Chemistry Fellowship.  The authors acknowledge the University of Chicago’s Research Computing Center for computing resources.

\end{twocolumngrid}

\bibliographystyle{unsrt}

\end{document}